\documentclass[twocolumn,showpacs,amsmath,amssymb,superscriptaddress,aps,pre]{revtex4-1}
\pdfoutput=1
\usepackage{graphicx, color} 
\usepackage{dcolumn}
\usepackage{bm,empheq}
\usepackage{units}
\usepackage{enumitem}
\usepackage{hyperref}

\newcommand{\ba}{\begin{eqnarray}}
\newcommand{\ea}{\end{eqnarray}}
\newcommand{\be}{\begin{equation}}
\newcommand{\ee}{\end{equation}}

  \newcommand{\xref}[1]{(\ref{#1})}
  \newcommand{\e}{{\text e}}
  
 \renewcommand{\d}{{\text{d}}}
  \newcommand{\D}{{\cal D}} 
  
\definecolor{mred}{rgb}{0.902, 0.0, 0.0} 
\definecolor{mblue}{rgb}{0.0, 0.0, 0.831}
\definecolor{mpurple}{rgb}{0.549, 0.0, 0.549}

\begin{document}
\title{
Three-phase coexistence with sequence partitioning in symmetric random block copolymers}
\author{Alice von der Heydt}
\email{heydt@theorie.physik.uni-goettingen.de}

\author{Marcus M\"uller}
\affiliation{Institut f\"ur Theoretische Physik, Georg-August-Universit\"at G\"ottingen, 
Friedrich-Hund-Platz 1, D-37077 G\"ottingen, Germany}
\author{Annette Zippelius}
\affiliation{Institut f\"ur Theoretische Physik, Georg-August-Universit\"at G\"ottingen,
Friedrich-Hund-Platz 1, D-37077 G\"ottingen, Germany}
\affiliation{Max-Planck-Institut f\"ur Dynamik und {Selbst}\-organisation, Bunsenstra{\ss}e~10, 
D-37073 G\"ottingen, Germany}

\date{\today}

\begin{abstract}
We inquire about the possible
coexistence of macroscopic and microstructured phases in random $Q$-block copolymers built of incompatible monomer types $A$ and $B$ with equal average concentrations.
In our microscopic model, one block comprises $M$ identical monomers.
The block-type sequence distribution is Markovian and characterized by the correlation $\lambda$.
Upon increasing the incompatibility $\chi$ (by decreasing temperature) in the disordered state,
the known ordered phases form:
for $\lambda > \lambda_c$, two coexisting macroscopic $A$- and $B$-rich phases, 
for $\lambda < \lambda_c$, a microstructured (lamellar) phase with wave number $k(\lambda)$. 
In addition, we find a fourth region in the $\lambda$-$\chi$ plane where these three phases coexist, 
with different, non-Markovian sequence distributions (fractionation).
Fractionation is revealed by our analytically derived multi-phase free energy, 
which explicitly accounts for the exchange of individual sequences between the coexisting phases.
The three-phase region is reached, either, 
from the macroscopic phases, via a third lamellar phase that is rich in alternating sequences, or, 
starting from the lamellar state, via two additional homogeneous, homopolymer-enriched phases. These incipient phases emerge with zero volume fraction.
The four regions of the phase diagram meet in a multicritical point $\left(\lambda_c, \chi_c\right)$, 
at which $A$-$B$ segregation vanishes. 
The analytical method, which for the lamellar phase assumes weak segregation, 
thus proves reliable particularly in the vicinity of $\left(\lambda_c, \chi_c\right)$. 
For random triblock copolymers, $Q=3$, we find the character of this point and the critical exponents 
to change substantially with the number $M$ of monomers per block. 
The results for $Q=3$ in the continuous-chain limit $M\to\infty$ 
are compared to numerical self-consistent field theory (SCFT), which is accurate at larger segregation.
 
\end{abstract}

\maketitle

\section{Introduction}

Random $A$-$B$ block copolymer melts 
represent an interesting class of materials both for applications, due
to their 
molecular self-organization for templating structures on the nanoscale as well as for everyday materials, 
and theoretically, as multi-component systems with 
competing interactions and a complex phase behavior \cite{bates-fred90rev,fred-miln91, fred92, bates-fred99}.

For copolymer mixtures, phase separation was first
addressed by Scott~\cite{scott52} within a mean-field theory of
multi-component demixing based on Flory-Huggins theory (see, e.g., \cite{flory53}). 
Scott computed the limits of stability of the disordered, mixed state against
macroscopic phase separation for arbitrary distributions of chain
composition (overall fraction of one monomer type). 
The coarse-grained description of ref.~\cite{scott52}, which
disregards the conformations of individual chains, was
subsequently extended by Bauer~\cite{bauer85} to assess the coexistence of multiple 
homogeneous phases and the equilibrium transition lines. The method was
applied to random copolymers by Nesarikar \textit{et al.}~\cite{nesarikar}, who
computed the phase diagram for various chain lengths and average 
compositions. The system is treated as a multi-component mixture, 
with components distinguished solely by composition. 
Upon increasing the incompatibility, 
successive separations into a growing number of homogeneous phases with different compositions are observed.

Taking into account the internal structure of the chains and the
block-type \emph{sequences\/} in a melt of random block copolymers is
crucial for the description of microstructured states (often termed microphase separation)
\cite{shak-gutin89,fred-miln91,fred92,sem99,subsem02}, see the example in the right panel of fig.~\ref{fig_triblocks}.   
Fredrickson, Milner, and Leibler \cite{fred-miln91,fred92} 
formulated a microscopic model for random block copolymers, 
with one block composed of $M$ identical monomers and with the block-type sequence distribution parameterized by a correlation $\lambda$.
Based on this model, they derived a mean-field free energy of Landau form in the limit of many blocks, $Q\to\infty$. 
The resulting phase diagram 
shows an isotropic Lifshitz point, separating a line of instabilities with zero wave number
(macroscopic phase separation) from a line of instabilities with finite
wave number (microphase separation), cf.\ the lines in fig.~\ref{fig-phasediag-sample}.

Several attempts have been made to go beyond mean-field theory and
to consider the effects of fluctuations, predicted to be
particularly important for the instability at finite wave number
\cite{brazov75,fred-helf87}. Whereas the early works \cite{sfatos-gutin-shak94_1,dob-erukhim95}
deduced complete stability of the disordered state against microphase separation, it
was later shown that proper inclusion of a local term in the
Landau-Wilson free energy restored microphase separation \cite{gutin-sfatos-shak94_2}. 
The transition was found to be weakly first order, yet wavelength and amplitude of
the microstructured phases matched the mean-field predictions \cite{fred92} rather well.


Monte Carlo simulations for symmetric random copolymer melts 
with different numbers $Q$ of blocks per chain were
performed by Houdayer and M\"uller~\cite{houd-mmuell02, mmueller04}.
In contrast to the mean-field calculations \cite{fred92}, 
macroscopic phase separation was found only for small $Q$ 
(in a $\lambda$-range shrinking with increasing $Q$), and
further increasing incompatibility in the two coexisting homogeneous phases 
resulted in a remixed state.
The latter was interpreted as the coexistence of three phases, two
homogeneous ones and a third microstructured one with symmetric
composition, as predicted for random diblock copolymers ($Q=2$), by simulation \cite{Muller96} and self-consistent field theory (SCFT) \cite{janert-schick97_2}. 
For $Q=3$, the simulations 
\cite{mmueller04} 
pointed to a three-phase coexistence 
with fractionation according to sequences: While the two homogeneous phases 
displayed a higher content of homopolymers, copolymers accumulated in the microstructured phase. 

In this paper we aim at an analytical theory for three-phase
coexistence due to \emph{sequence-specific fractionation\/}: According
to its internal structure, in particular the number of bonded $A$-$B$ contacts, a
sequence class, e.g.\ $AAB$/$BBA$, may have different concentrations in homogeneous and
structured phases. Our global copolymer distributions are 
symmetric in A/B content, which causes the $A$-rich and $B$-rich phases
in a macroscopically separated state to map onto each other by permutation of $A$ and B.
The distributions of these two phases, though different in composition,  
are not called fractionated, 
since they preserve the global concentration of a \emph{sequence\/} class, e.g., of $AAB$/$BBA$.
Their $A$ excesses of opposite signs result from exchange of $A$- and $B$-rich subspecies 
only within one sequence class. The $A$-rich phase, for instance, successively substitutes $BBA$ chains with $AAB$ chains, inversely the $B$-rich phase. 
In contrast, we define sequence-specific fractionation 
to alter the sequence (class)
concentrations in parts of the system such that microphase separation is
favored in one part, while macrophase separation persists in the other.

Our main results are the phase diagrams as a function of block correlation $\lambda$ and incompatibility $\chi$
(see, e.g., fig.~\ref{fig-chi2m-chih-M3} below)
showing a three-phase coexistence region of two homogeneous and one lamellar phase. 
Additional information concerns the volume fractions, the wavelengths, and the sequence distributions of the fractionated states, as well as the behavior at the multicritical point.
Some results provided by the analytical method have been briefly presented in ref.~\cite{hzm-mamo10}.
Coming from the macroscopically phase-separated state, a
lamellar phase emerges with zero volume fraction (called shadow) 
and with finite amplitude; 
similarly, coming from the lamellar state, two additional homogeneous phases appear as shadows. 
The nature of the multicritical point, where four 
states of the system meet, depends on the number $M$ of monomers per block: 
For $M<7$, the wave number of the incipient lamellar phase vanishes continuously
on approach to the multicritical point, and the segregation amplitude vanishes linearly. 
For $M\geq 7$ and particularly in the limit of continuous chains, the wave number remains finite,
giving rise to metastable regions on both sides. In this case, the critical exponent for the segregation amplitude is $0.5$.
Detailed sequence-concentration diagrams of the coexisting phases show the partitioning according to their morphologies. Except for at the multicritical point itself, the shadow phase emerges with a finite deviation from the global, $\lambda$-defined distribution.
A numerical SCFT study for continuous triblock copolymers 
covers larger segregation amplitudes, but yields 
a similar phase behavior. 

The paper is organized as follows: 
The microscopic model is introduced in sec.~\ref{chap-model}. 
Free energies of macroscopic and microstructured phase separation
and the sequence-specific correlators are derived in sec.~\ref{chap-F}. 
In sec.~\ref{chap-f-frac}, 
we construct the free energy of a fractionated state and discuss the resulting phase diagrams in
sec.~\ref{chap-phase-diag}. 
SCFT as a 
complementary approach is presented in sec.~\ref{chap_SCFT}. 
Sequence fractionation is addressed in sec.~\ref{sec-fract-distr}.
A discussion of the methods 
is given in sec.~\ref{chap-discussion}, 
followed by conclusions and an outlook in sec.~\ref{chap-conclusion}.

\section{\label{chap-model}Model}

\subsection{\label{sec-def-tribl}Symmetric random block copolymers}

\begin{figure}[h]
\includegraphics[width=3.5cm]{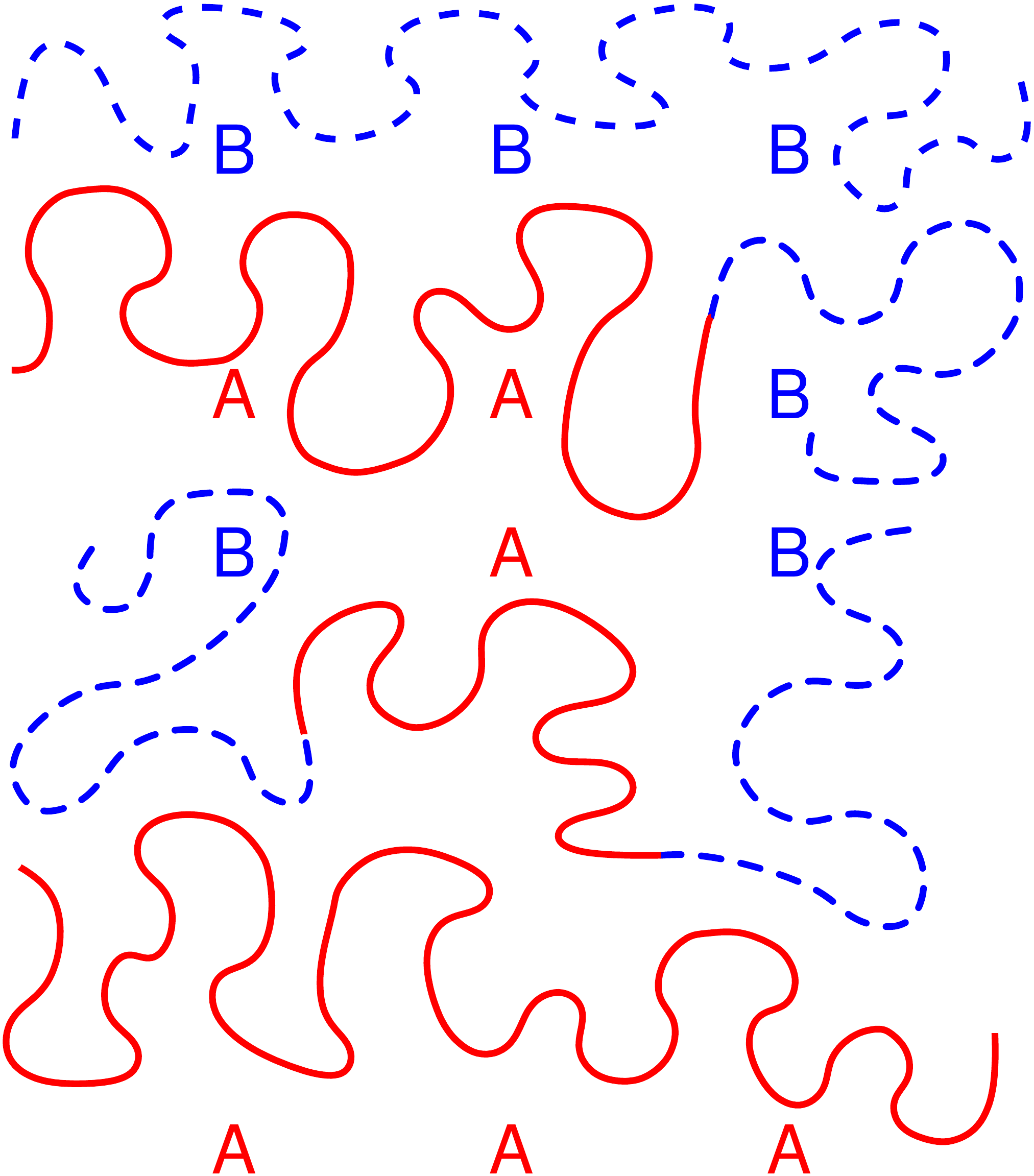}
\includegraphics[width=3.5cm]{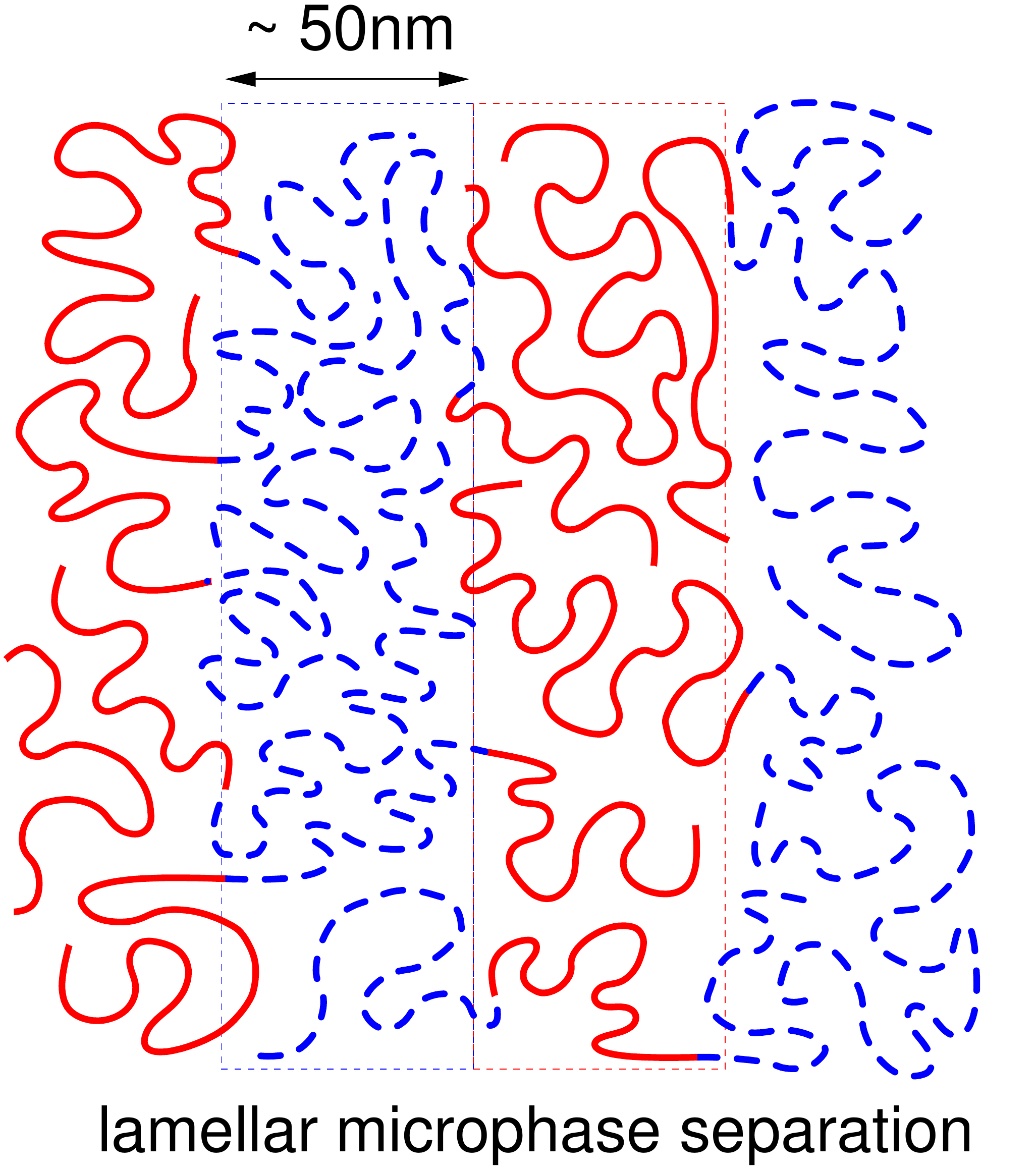}
\caption{(Color online) Cartoon of triblock copolymers and lamellar phase separation.\label{fig_triblocks}}
\end{figure}

We consider an incompressible melt of $j=1,\ldots,N$ linear, random $A$-$B$ block
copolymers in a volume $\tilde V$ (fig.~\ref{fig_triblocks} shows triblocks). All chains have 
degree of polymerization $L = QM$, each of the $Q$ blocks
comprises $M$ identical monomers,
\be
          \overset{\text{block } 1}{
            \underbracket{{\color{mred}\bullet\bullet\bullet\bullet\bullet}}_{M}}\,
          \overset{\text{block } 2}{
            \underbracket{\color{mblue}\circ\circ\circ\circ\circ}_{M}}\,
            \overset{\ldots}{\text{---}}\,
          \overset{\text{block } Q}{
            \underbracket{\color{mred}\bullet\bullet\bullet\bullet\bullet}_{M}}.
\ee
Both types of monomers or segments are assumed to have the same
statistical length, $b$. To formulate the effective repulsive interaction
between monomers of different types (see section \ref{sec-pot}),
we introduce an $A$ excess variable $q_j(s)$ for the type
of segment $s$ on polymer $j$, which takes the values $+1$ for $A$ and $-1$ for $B$. 
The type sequences of symmetric random block copolymers are 
generated by a Markovian polymerization process with average $A$ excess $q = 2p-1 = 0$
($p$ is the global concentration of type-A segments)
and block-type correlation 
\be 
\lambda \mathrel{\mathop:}= (1 - 2 p_{AB}) \in\left[-1, +1\right]
\label{def-lambda}\ee
of adjacent blocks along a chain \cite{fred92}. 
Here, $p_{AB}$ denotes the probability that a block of type A
is attached to one of type $B$ in the synthesis.
Assuming homogeneity, 
$p$ and $\lambda$ are independent of the position on a chain.
Positive $\lambda$ signal a preference for homopolymers, 
and $\lambda = 0$ describes ideal (uncorrelated) block sequences. 
This model for the synthesis 
amounts to choosing the simplest nontrivial distribution 
with only two parameters, $p$ and $\lambda$.
With $p=\nicefrac{1}{2}$ in our case, 
the corresponding transition matrix $\hat M$ 
for the probability vector $[p_{A}(\beta), 1- p_{A}(\beta)]^T$
(probabilities to find $A$, respectively $B$, at block $\beta$) reads
\be \label{t-matrix}
{ \hat M} = \left(\begin{array}{cc} 
\frac{1 + \lambda}{2} 	& \frac{1 - \lambda}{2}\\
\frac{1 - \lambda}{2} 		& \frac{1 + \lambda}{2} 
\end{array}\right).
\ee 
Its diagonalized form is used to compute 
the probabilities of individual sequences in the $\lambda$-distribution, 
and moments of the $A$ excess distribution.
Once generated, the block 
sequences remain fixed, i.e.,
thermal averaging affects only the chains' center of mass locations and their conformations. 
For a finite number of different sequences, 
a concentration for each sequence is well-defined in the thermodynamic limit. 
Hence, for finite $Q$, the quenched disorder due to the fixed block types on one chain
can be effectively translated to a multi-component system. 
 
A chain can contain 
$0$ to $Q$ blocks of type $A$,
which defines $Q + 1$ classes of chains. 
This classification in the ``crushed polymer approximation'' (see, e.g.,
\cite{bauer85}, \cite{nesarikar}) is sufficient to study the separation into homogeneous phases (see sec.~\ref{sec-hom-phases} below).
However, it neglects differences in the \emph{sequence\/} of the blocks, 
i.e., the spatial structure of the chains (for example, the average $A$ excess for both
$AAB$ and $ABA$ chains is $\nicefrac{1}{3}$). 
The spatial correlation of types
along a chain is essential for the formation of structured phases
with nonzero wave numbers.

\subsection{\label{sec-pot}Potentials}

The Hamiltonian ${\cal {H}}$ consists of three parts,
\begin{equation}
\label{Hamiltonian}
{\cal {H}}={\cal{H}}_{\text W}+{\cal{H}}_{\kappa}+{\cal{H}}_{\chi} ,
\end{equation}
which reflect intra- and interchain 
interactions of the monomers on a mesoscopic 
level. Explicitly, for the former we consider the connectivity 
of Gaussian or ideal polymers \cite{flory69} acting between monomers on the same chain, 
for the latter excluded volume 
and incompatibility acting between all monomers,
in units of $k_{\rm B} T$:
\begin{subequations}\label{H_total}
\begin{align} 
{\cal{H}}_{\text W} &= \frac{1}{4}\sum_{j=1}^N\sum_{s=1}^{L-1} 
\Bigl( \bm r_j(s+1) - \bm r_j(s) \Bigr)^2, \label{H_wiener}\\
{\cal{H}}_{\kappa} &= 
\frac{\kappa}{2\rho_0}\!
\underset{(j_1,s_1)\neq(j_2,s_2)}{
\sum_{j_{1,2}=1}^N\sum_{s_{1,2}=1}^L
}
\!U \Bigl( |\bm r_{j_1}(s_1) - \bm r_{j_2}(s_2)| \Bigr), \label{H_ev}\\
{\cal{H}}_{\chi} &=\label{H_chi} 
-\frac{\chi}{4\rho_0} \sideset{}{'}\sum_{j_{1,2}, s_{1,2}}\!
q_{j_1}(s_1)q_{j_2}(s_2) 
W \Bigl( |\bm r_{j_1}(s_1) - \bm r_{j_2}(s_2)| \Bigr),
\end{align}
\end{subequations}
where the primed sum in eq.~\xref{H_chi} is shorthand for the constrained sum in eq.~\xref{H_ev}
(the constraint can be dropped in the thermodynamic limit).
Spatial variables $\bm r$ are dimensionless, rescaled from
physical positions $\bm R$ via
\be r_{\alpha} = \frac{\sqrt{2d}R_{\alpha}}{b},\quad \alpha = 1,2,\ldots, d\ee
with $b$ the 
rms end-to-end distance of a Kuhn statistical segment
and $d$ the spatial dimension.
Accordingly,
the  constant dimensionless monomer number density is 
\be \rho_0\mathrel{\mathop:}= \frac{NL}{V} \mathrel{\mathop:}= \frac{N L b^d}{\tilde V (2d) ^{d/2}}.\label{def_rho_0}\ee
One effective segment of our model usually represents many physical monomeric repeat units, 
as to fulfill the prerequisite of statistical independence of subsequent bond vectors 
in the coarse-grained Gaussian chain model. 

The excluded volume interaction eq.~\xref{H_ev} must be accounted for, even if we later perform  the incompressible limit, since $A$ excess and total density fluctuations are coupled. 
The pair potentials $U(r)$, $W(r)$ are supposed to be short-ranged, 
and we approximate them by $\delta$ functions,
neglecting short-wavelength fluctuations.
Conceptually, Gaussian chain connectivity and compressibility are 
effective potentials, which are obtained after integrating-out microscopic degrees of freedom.
They are chiefly of entropic origin and thus originally proportional to $k_{\rm B}T$. 
The Flory parameter $\chi$ expresses in an empirical way the
local free-energy change per monomer 
due to $A$-$B$ contacts 
compared to a surrounding of monomers of the same type 
with larger attractive potentials
\cite{flory53}. 
Its main part is usually enthalpic,
such that in the normalized eqs.~\xref{H_total}, $\chi$
is inversely proportional to temperature, 
$\chi\propto T^{-1}$, and 
increasing incompatibility is equivalent to cooling. 
In the following, $k_{\rm B}T$ is set to unity. 

\subsection{\label{sec-orderpar}Order parameter}

A convenient order parameter that detects separation into $A$- and $B$-rich domains (phases) 
is the thermal average of the local excess of $A$ segments  \cite{fred92}, 
\be 
\sigma(\bm r) = \sum_{j=1}^N\sum_{s=1}^L q_j(s) \delta\left(\bm r
  - \bm r_j(s)\right) = \varrho_{A}(\bm r) - \varrho_{B}(\bm r),
\label{sigma-def}\ee
i.e., the difference of segment densities due to $A$ and B.
As a second field, we introduce the total segment density
\be 
\varrho(\bm r) = \sum_{j=1}^N\sum_{s=1}^L \delta\left(\bm r - \bm
  r_j(s)\right) =  \varrho_{A}(\bm r) + \varrho_{B}(\bm r).
\label{rho-def}\ee
With these fields, and in the limit $W(r) \to \delta(r) $, the incompatibility \xref{H_chi} takes the standard form \cite{matsen02b}
\ba
{\cal{H}}_{\chi} 
&=&\frac{\chi}{\varrho_0}\int\d^d r \,
\left(\varrho_{A} (\bm r) - \frac{\varrho(\bm r)}{2}\right)
\left( \varrho_{B}(\bm r) - \frac{\varrho(\bm r)}{2} \right) \nonumber\\
&=& - \frac{\chi}{4 \varrho_0}\int\d^d r\, \Bigl( \sigma (\bm r) \Bigr)^2.
\ea
Note that as a zero of the incompatibility energy 
we have chosen the homogeneously mixed state where the local densities 
of $A$ and $B$ 
coincide with their global fractions throughout the system.
Analogously, the excluded volume interaction eq.~\xref{H_ev} in the limit $U(r) \to \delta(r)$ is
\be
{\cal H}_{\kappa} = \frac{\kappa}{2\varrho_0}\int\d^d r\, \Bigl( \varrho(\bm r) 
\Bigr)^2. \ee

\section{\label{chap-F}Free energy}

\begin{figure}[h!]
\includegraphics[width=7.3cm]{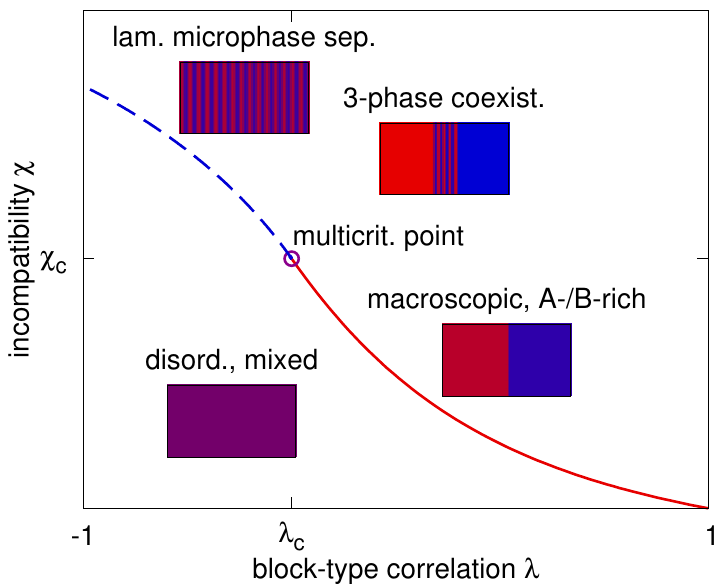}
\caption{(Color online) Qualitative phase diagram 
of random block copolymers. 
Global instabilities of disordered melt:
solid (red) line: macroscopic phase separation for $\lambda > \lambda_c$, 
dashed (blue) line: lamellar phase for $\lambda < \lambda_c$. 
Fractionation creates an in-between state with three coexisting phases. 
\label{fig-phasediag-sample}}
\end{figure}

In order to assess the phase diagrams, particularly phase coexistence
for random block copolymer melts, we compute the free energies of basic phase-separated states.
The two important control parameters are the incompatibility $\chi$ and the block-type
correlation $\lambda$. Figure~\ref{fig-phasediag-sample} shows the topology of
the phase diagrams we will derive below. 
As discussed already by Leibler and co-workers~\cite{leibler80,fred92}, 
the  disordered state of the symmetric melt becomes unstable toward either
macroscopic or lamellar phase separation, depending on $\lambda$. 
Between these two well-known states, a new
state will be shown to become stable, viz.\ the
coexistence of three phases: an $A$-rich one, a $B$-rich one, both homogeneous, and a
lamellar phase. Coming from the macroscopically phase-separated state,
the new phase is created by expulsion of chains with many
$A$-$B$ contacts from the homogenous cloud phases 
(for the terms ``cloud'' and ``shadow'' phase, see \cite{sollich02}) into a subsystem, 
which then displays deviations from the global $\lambda$-distribution. 
This fractionation increases the $A$ excess amplitude in the homogeneous phases. 
More explicitly, lamellae can appear in the new phase because the altered
sequence distribution with fewer homopolymers gives rise to a maximum of
the structure factor at nonzero wave number, whereas the structure
factor of the global distribution favors macroscopic phase
separation. Conversely, starting from the lamellar
phase, homopolymer chains are expelled into two new homogeneous
phases. Thereby, 
for values of $\lambda$, at which the global sequence distribution favors lamellae,
homogeneous phases become stable in a subsystem, resulting in a fractionated state.

First, we present the free-energy densities of homogeneous phases and of
lamellae separately. The expressions are deliberately kept as simple as possible
to focus on the effect of varying sequence concentrations. 
In the following section, 
we go on to set up a fractionated multi-phase free energy, 
allowing for sequence distributions different from the global one, 
and discuss three-phase coexistence.

\subsection{Free energy functional\label{sec:fmps}}
Starting from the Hamiltonian of eq.~\xref{Hamiltonian}, 
we aim at computing the canonical partition function 
\begin{equation}
{\cal{Z}}=\int \prod_{j,s}\d{\bm r}_j(s) \e^{\textstyle -\cal{H}}.
\end{equation}
Pair 
interactions 
are formally decoupled 
via functional integrations over the collective density fields $\sigma$ and $\varrho$,
and over two conjugated interaction fields $\hat\sigma$ and $\hat\varrho$ 
(with Fourier modes $\left\{ \hat\sigma_{\bm k}, \hat\varrho_{\bm k} \right\}$)
that restrict $\sigma$ to the $A$ excess and $\varrho$ to the total segment density 
(cf.~eqs.~\xref{sigma-def} and \xref{rho-def}): 
\begin{align}
{\cal{Z}} = \int\D & \left[ \hat\sigma,\sigma, \hat\varrho,\varrho \right]  
\label{part-fct}\\
\exp\biggl\{ 
&\sum\limits_{\bm k \neq 0} %
\Bigl(
\frac{\chi}{4 NL} \sigma_{\bm k}\sigma_{-\bm k}
- \frac{\kappa}{2 NL} \varrho_{\bm k}\varrho_{-\bm k}
\Bigr.\biggr.
\nonumber\\
\biggl.
& \Bigl.
\mbox{} + i \hat\sigma_{\bm k} \sigma_{-\bm k}
+ i \hat\varrho_{\bm k} \varrho_{-\bm k}
\Bigr)
+ \sum_{\nu} N_{\nu} \ln \hat z_{\nu}[\hat\sigma,\hat\varrho]
\biggr\}. 
\nonumber
\end{align}
In this expression, the inner conformational integrations 
have factorized into 
single-chain partition functions $\hat z_j$. 
All $N_{\nu}$ chains with a given block-type sequence $\nu$,
which is characterized by the segment types $\left\{q_{\nu} (s) \right\}$,
contribute 
the single-sequence partition function
\begin{align}
\lefteqn{
\hat z_{\nu} \left[\hat\sigma,\hat\varrho\right] \mathrel{\mathop:}= 
}\label{q-sc}\\
&\left\langle
\exp\biggl\{
- i \sum\limits_{\bm k\neq 0} \sum_{s} 
\bigl(
\hat\sigma_{\bm k} q_{\nu}(s) + \hat\varrho_{\bm k}
\bigr)
\e^{-i\bm k\cdot\bm r(s)}
\biggr\}
\right\rangle.
\nonumber
\end{align} 
Here, $\left\langle (\ldots) \right\rangle$
denotes the conformational average
\be\label{wiener-av}
\frac{
\int\D \bm r(s)(\ldots)\exp
\left\{
- \frac{1}{4}\sum_{s=1}^{L-1}
\bigl( \bm r(s+1) - \bm r(s) \bigr)^2
\right\}
}{
\int\D \bm r(s)\exp
\left\{
- \frac{1}{4}\sum_{s=1}^{L-1}
\bigl( \bm r(s+1) - \bm r(s) \bigr)^2
\right\}
}
\ee
for one Gaussian chain (cf.\ eq.~\xref{H_wiener}).
Combinatorial prefactors $1/N_{\nu}!$,
homogeneous contributions ($\bm k = 0$), 
and the conformational partition functions of noninteracting Gaussian chains
have been divided out in eq.~\xref{part-fct}, 
since we are interested in the free energy of a global ordered state \emph{relative\/} to the disordered, homogeneous state. 

In order to perform the saddle-point approximation,
we choose to first integrate out the amplitudes of the physical fields 
in favor of the conjugated ones,
contrasting with the {\itshape procedere} in, e.g., refs.~\cite{leibler80,fred92}
(see the note below eq.~\xref{f_4th}). 
From eq.~\xref{part-fct}, we obtain the linear relations
\be 
\sigma_{\bm k} = - i \frac{2 NL}{\chi} \hat\sigma_{\bm k}
\mbox{ and } 
\varrho_{\bm k} = i \frac{NL}{\kappa} \hat\varrho_{\bm k},
\label{sp1}\ee
and, for convenience, rescale the conjugated fields as
\be
\hat\sigma_{\bm k} \mathrel{\mathop:}= i \frac{\chi}{2 NL} \hat\tau_{\bm k}
\mbox{ and }
\hat\varrho_{\bm k} \mathrel{\mathop:}= \frac{-i}{NL} \hat\omega_{\bm k}
\label{sp2}\ee
before insertion into $\cal{Z}$.
The resulting partition function in saddle-point approximation is 
\be
\label{part-fct-sp} 
{\cal{Z}}  \approx
\int\D[\hat\tau, \hat\omega]\,
\exp\Bigl\{
- N h [\hat\tau, \hat\omega]
\Bigr\},
\ee
with the effective Hamiltonian (per chain)
\begin{align}\lefteqn{
h \left[\hat\tau,\hat\omega \right]
=}\label{eff-h-sp}\\
& \frac{1}{4N^2L}
\sum_{\bm k\neq 0}\left(
	\chi  \hat\tau_{\bm k}\hat\tau_{-\bm k}
     	- \frac{2}{ \kappa} \hat\omega_{\bm k}\hat\omega_{-\bm k} 
\right)-\sum_{\nu} p_{\nu}\ln z_{\nu}\left[\hat\tau,\hat\omega \right],
\nonumber
\end{align}
and the single-sequence partition functions
\begin{align}
\lefteqn{ z_{\nu} \left[\hat\tau,\hat\omega \right] = }\\
&\left\langle
\exp
\biggl\{
\frac{1}{2NL}
\sum_{\bm
  k\neq 0}\sum_{s}\bigl(\chi q_{\nu}(s)\hat\tau_{\bm k} - 2 \hat\omega_{\bm k}\bigr)
\e^{-i\bm k\cdot\bm r(s)} \biggr\}
\right\rangle.
\nonumber\end{align}
The probabilities $p_{\nu} \mathrel{\mathop:=} N_{\nu}/N$ define, 
in the thermodynamic limit, the sequence distribution over the up to $2^Q$ 
possible realizations 
of a random, binary $Q$-block copolymer. 
(For $Q\geq 2$, the actual number of different sequences is smaller 
due to the symmetry with respect to the two ends,
see below for triblock copolymers.)

Anticipating small field amplitudes,
the next step is to expand the effective Hamiltonian $h$ eq.~\xref{eff-h-sp} 
into a series in both fields. 
Restricting ourselves to systems with global $A$-$B$ symmetry,
the expansion contains no terms of odd order in $\hat\tau$ (the field conjugated to the $A$ excess),
since for $n$ odd, moments of the $A$ excess distribution,
\be \label{nth-mom}
m_n 
\mathrel{\mathop:}=
\frac{1}{L^{n}} \sum_{\nu} p_{\nu}\!
\sum_{s_1,\ldots, s_n} q_{\nu}(s_1)\ldots q_{\nu}(s_n),
\, n 
\in {\mathbb N},
\ee
are zero.
A sufficiently large compression modulus $\kappa$ will prevent 
instabilities with respect to fluctuations of the total density. 
Hence, we can 
eliminate 
their conjugated amplitudes $\hat\omega_{\bm k}$  
perturbatively in favor of $\hat\tau_{\bm k}$
and obtain, to lowest order, a quadratic dependence 
\begin{align} 
\lefteqn{ 
\hat\omega_{\bm k} =  }\label{rho_sp}\\
&\frac{\chi^2}{8 NL} \sum_{\bm k_1 \notin\{ \bm 0, \bm k\}}
\frac{ S^{(\alpha)}(\bm k_1, \bm k - \bm k_1) }{
\frac{2L}{\kappa} +  D \left(L, k^2\right)
}
\hat\tau_{\bm k_1}\hat\tau_{\bm k - \bm k_1} 
+ {\cal{O}}\left(|\hat\tau_{\bm k}|^4\right)
\nonumber
\end{align}
(see appendix \ref{app_wiener-av} for conformational averages of exponentials 
and appendix \ref{app_vertexfcts} for the correlators $S^{(\alpha)}$ and $D$).
Substituting back
this relation, and in the incompressible limit, $\kappa \to \infty$,
the consistent expansion up to fourth order in $\hat\tau$ 
yields
the free-energy functional per chain,
\begin{widetext}
\begin{align}
f \left[\hat\tau \right] 
=&\frac{\chi}{4N^2 L}
\sideset{}{'}\sum_{\bm k} \left(1 - \chi \frac{S(k^2)}{2L}\right) 
\hat\tau_{\bm k}\hat\tau_{-\bm k} 
\label{f_4th}\\
& \mbox{} + \frac{\chi^4}{384 (NL)^4}
\sideset{}{'}\sum_{
\begin{array}{c}
\scriptstyle
\bm k_1, \bm k_2, \bm k_3\\
\scriptstyle\bm k_1 +\bm k_2 + \bm k_3 \neq 0
\end{array}
}
\left\{
3 \frac{S^{(\alpha)}(\bm k_1, \bm k_2) S^{(\alpha)}(\bm k_3, - \bm k_1 - \bm k_2 - \bm k_3)}{
D \left(L, (\bm k_1+\bm k_2)^2 \right)}
- S^{(\beta)}(\bm k_1,\bm k_2, \bm k_3)
\right\} 
\hat\tau_{\bm k_1} \hat\tau_{\bm k_2} \hat\tau_{\bm k_3} \hat\tau_{- \bm k_1 - \bm k_2 - \bm k_3} \nonumber\\
&\mbox{} + \frac{\chi^4}{128 (NL)^4}
\sideset{}{'}\sum_{\bm k_1, \bm k_2} 
\left\{
S^{(\gamma)}(k_1^2, k_2^2) - S(k_1^2) S(k_2^2)
\right\}
\hat\tau_{\bm k_1} \hat\tau_{-\bm k_1} \hat\tau_{\bm k_2} \hat\tau_{-\bm k_2}
+ {\cal{O}}\left(|\hat\tau_{\bm k}|^6\right),
\nonumber
\end{align}
\end{widetext}
with $\sum_{\bm k}^{'}(\ldots)\mathrel{\mathop:}=\sum_{\bm k \neq 0}$. 
The global second-order 
correlator $S(k^2)$, 
called structure factor in the following and discussed in sec.~\ref{sec-S-k-lambda}, 
is given by
\begin{align} 
S(k^2) 
&= 
\sum_{\nu} p_{\nu}\, \sum_{s_1,s_2=1}^L q_{\nu}(s_1)q_{\nu}(s_2) \e^{-k^2|s_2-s_1|}
\label{S-p-nu}\\&=\mathrel{\mathop:} 
\sum_{\nu} p_{\nu}\, S_{\nu}(k^2), 
\nonumber
\end{align}
written as an average over intra-chain correlators $S_{\nu}(k^2)$ of single block-type sequences. These and the correlators $S^{(\alpha)}$, $S^{(\beta)}$, and $S^{(\gamma)}$ are defined in appendix \ref{app_vertexfcts}.
In our global sequence distribution, 
the probabilities $p_{\nu}$ will be confined to $\lambda$-defined values 
(see eqs.~\xref{p_lambda} below), 
but can take arbitrary values in a fractionated subsystem.

As suggested by the functional eq.~\xref{f_4th}, 
we assign the conjugated field $\hat\tau$ the r\^ole of the order parameter,
since at the saddle point level, to which we adhere, 
averages of $\hat\tau$ and $\sigma$ are identical (cf.~eqs.~\xref{sp1} and \xref{sp2}).
However, correlations of the conjugated field
are not proportional to those of the field itself, cf., e.g., \cite{reister-mm-binder01}. 
Therefore the vertices in eq.~\xref{f_4th} differ from those of the functional 
of $\sigma$ in refs.~\cite{leibler80,fred92} (apart from differences due to restrictions, 
e.g., to continuous chains with many blocks, which we do not impose).
For instance, second moments 
of the amplitudes of $\sigma$
can be recovered from those of $\hat\tau$ via
\begin{align} 
&\left\langle \sigma_{\bm k_1} \sigma_{- \bm k_2}\right\rangle_{\cal{H}}
  - \left\langle \sigma_{\bm k_1} \right\rangle_{\cal{H}} 
    \left\langle \sigma_{-\bm k_2} \right\rangle_{\cal{H}}
=\nonumber\\
&\left\langle \hat\tau_{\bm k_1} \hat\tau_{- \bm k_2}\right\rangle_{\cal{H}}
  - \left\langle \hat\tau_{\bm k_1} \right\rangle_{\cal{H}} 
    \left\langle \hat\tau_{-\bm k_2} \right\rangle_{\cal{H}}
- \frac{2NL}{\chi} \,\delta_{\bm k_1, - \bm k_2},
\label{var-conj-field}\end{align}
where $\left\langle\cdot\right\rangle_{\cal{H}}$ is the canonical average, 
eqs.~\xref{part-fct}, respectively eq.~\xref{part-fct-sp}.   

Aiming first at the simplest description, and in the spirit of a Landau free energy, 
we ignore the wave-vector dependence of the fourth-order coefficients in eq.~\xref{f_4th}, 
i.e., we evaluate the correlators in the limit $\bm k_{\text r} \to \bm 0$, 
(in secs.~\ref{sec-landau-k4} and \ref{sec-frac-compl}, we will relax this approximation):
\begin{align}
&f_0 \left[ \hat\tau \right] 
= \label{f_4thk0} 
 \frac{\chi}{4N^2L}   
 \sideset{}{'}\sum\limits_{\bm k} 
\left(1 - \chi \frac{S(k^2)}{2L}\right) \hat\tau_{\bm k}\hat\tau_{-\bm k} 
\\  & \mbox{} +  \frac{\chi^4}{128 N^4} 
    \left\{  \left( m_2^2-\frac{m_4}{3} \right)
   \sideset{}{'}\sum_{\bm k_1,\bm k_2,\bm k_3}
   \hat\tau_{\bm k_1}\hat\tau_{\bm k_2}\hat\tau_{\bm k_3}
   \hat\tau_{-\bm k_1-\bm k_2 - \bm k_3} 
   \right.\nonumber\\  & \left. \mbox{} + 
   \left( m_4 - m_2^2 \right)
   \sideset{}{'}\sum_{\bm k_1,\bm k_2}
   \hat\tau_{\bm k_1}\hat\tau_{-\bm k_1} \hat\tau_{\bm k_2}\hat\tau_{-\bm k_2}
   \right\},
\nonumber
\end{align} 
with the 
moments $m_2$, $m_4$ from eq.~\xref{nth-mom}.

\subsection{\label{sec-S-k-lambda}Structure factor and multicritical point}

The second-order structure factor $S\left( k^2\right)$ for a distribution of
sequences sets the limits of stability of
the homogeneously mixed melt. For our global Markovian distributions, 
solely the correlation parameter $\lambda$ decides whether
the maximum position of $S\left(k^2\right)$ is located at zero
or at finite wave number. In the former case, the disordered state 
becomes unstable with respect to macroscopic phase separation, in
the latter case to microphase separation \cite{fred-miln91}. 
Upon decreasing $\lambda$, the maximum position of  $S\left(k^2\right)$ becomes nonzero
at a critical correlation $\lambda_c(M)$, depending on the number
$M$ of segments per block. 
The corresponding point in the $\lambda$-$\chi$ plane where
the lines of macroscopic, respectively lamellar phase separations meet is
termed a multicritical point, since also the transition lines to
three-phase coexistence must end here.

For a $\lambda$-distribution of $Q$-blocks with finite $M$, the global 
$S\left(k^2\right)$ can be calculated 
from the probabilities of all type combinations of two segments with a given intrachain 
distance (in blocks)
using the transition matrix $\hat M$ 
(cf.\ eq.~\xref{q-corr-lambda} in appendix \ref{app_vertexfcts}): 
\ba\label{S-k-nat}
\lefteqn{
S\left(k^2\right) 
= Q D\left( M, k^2\right) }\\ 
&& \mbox{} 
+ \frac{2 \lambda \e^{-Mk^2} \sinh^2\left(\frac{Mk^2}{2}\right)}{\left(1 - \lambda
  \e^{-Mk^2}\right)\sinh^2\left(\frac{k^2}{2}\right)}\left[Q - \frac{1
- \left(\lambda \e^{-Mk^2}\right)^Q}{1 - \lambda\e^{-Mk^2}}\right]
\nonumber
\ea
with the dimensionless wave number 
$k^2\mathrel{\mathop:}= b^2 \tilde k^2/(2d)$ and $\tilde k$ the physical wave number.
The discrete Debye function $D\left( L, k^2 \right)$ is given in eq.~\xref{debye-L}.

In the following, we restrict ourselves to the case of symmetric random triblock copolymers, $Q=3$.  
This system features six different species, which we group into 
only three different (classes of) sequences,
\begin{subequations}
\label{species-def}
\begin{align}
\mbox{homopolymers: } & LLL;    \label{spec1}
\\ 
\mbox{copolymers: } & KLL;  \label{spec2}
\\ 
& LKL, \label{spec3}
\end{align}
\end{subequations}
$K, L \in \left\{A, B\right\}$, $K \neq L$,
according to 
unfavorable intrachain $A$-$B$ contacts.
Generally,
pairs of species like $AAB$ and $BBA$ are related by blockwise $A$-$B$ permutation and have the same topology of intrachain $A$-$B$ contacts, and thus the same structure factor.
To label these sequences, 
the index $1$ is assigned to homopolymer chains \xref{spec1}, 
$2$ to copolymer chains with two 
adjacent blocks of the same type \xref{spec2}, 
and $3$ to  strictly alternating chains \xref{spec3}.
For a $\lambda$-distribution, 
the sequence (class) concentrations are
\begin{subequations}
\label{p_lambda} 
\ba 
p_1(\lambda)&=&\frac{(1+\lambda)^2}{4},
\\
p_2(\lambda)&=& \frac{1-\lambda^2}{2},
\\
p_3(\lambda)&=&\frac{(1-\lambda)^2}{4}.
\ea
\end{subequations}

At a critical correlation $\lambda_c(M)$, 
we find the following transition from macroscopic to lamellar phase separation: 
\begin{enumerate}[label = \emph{\alph*})]
\item\label{lifshitz}
For $M \leq 6$, the maximum position of $S\left( k^2 \right)$ is 
at $k_0 = 0$ for all $\lambda \geq \lambda_c(M)$) and grows 
\textit{continuously\/} from $k_{0}=0$
when $\lambda$ falls below $\lambda_c(M)$ (see fig.~\ref{fig-chi2m-chih-M3} below). 
The critical value of the correlation, $\lambda_c(M)$, is reached when the second derivative of 
$S\left( k^2 \right)$ at $k=0$ changes sign:
\be
\lambda_c(M) = - \frac{1}{2}\left(1 - \frac{1}{M}\right),\quad M\leq 6 
\label{eq-lambda_c}
\ee

\item\label{discont}
For $M>6$, however, a \emph{second maximum\/} of $S\left( k^2 \right)$ at $k>0$
evolves already for $\lambda > \lambda_c(M)$ (see fig.~\ref{fig-chi2m-chih-cont}). Now, the
critical value $\lambda_c$ is the one at which the second
maximum (associated with a metastable lamellar phase) 
attains a higher value than the one at $k=0$, and is accessible numerically only.
\end{enumerate}

For continuous Gaussian triblocks (segments indexed by a contour parameter instead of an integer) with unaltered coil diameter, the
structure factors are computed in the combined limit $M\to\infty$,
$b^2\to 0$, $M b^2=\text{const}$, abbreviated as $\lim_{M\to\infty}$, 
preserving the finite number of blocks, here $Q=3$, and the rms end-to-end distance 
$R_{\text{block}} = \sqrt{M} b$. 
In this case, the wave number is conveniently rescaled with $R_{\text{block}}$.
For a $\lambda$-distribution of continuous triblocks, the global structure factor is  
\ba
\lefteqn{s\left( k^2 \right)\mathrel{\mathop:}=\lim_{M\to\infty} 
S\left( k^2 / M \right)/M^2
=}\label{s-k-nat}\\
&&3 g_{\text{D}}(1, k^2)
+ \frac{2\lambda\e^{-k^2} \sinh^2\left(\frac{k^2}{2}\right)}{
\left(1 - \lambda\e^{-k^2}\right) k^4/4}
\left[3 - \frac{1
- \left(\lambda \e^{-k^2}\right)^3}{1 - \lambda\e^{-k^2}}\right],
\nonumber
\ea
now with $k^2\mathrel{\mathop:} = R_{\text{block}}^2 \tilde k^2/(2d)$, 
and with the continuous Debye function
\be\label{debye_cont}  
g_{\text{D}}(\ell, k^2) \mathrel{\mathop:}= \frac{\e^{-\ell k^2} - 1 + \ell k^2}{k^4/2}.
\ee
Continuous triblocks realize case \ref{discont}, 
consistent with the case of triblocks with $M>6$ discrete segments. 
The wave number of the global ordered (lamellar) state, 
$k_0(\lambda)$, jumps \textit{discontinuously\/} to zero as $\lambda$
approaches $\lambda_c = -0.464$ from below. The lamellar phase
persists as a \textit{metastable\/} state for $\lambda > \lambda_c$, 
as well as macroscopic phase separation for $\lambda < \lambda_c$.
Remarkably, we discover this discontinuity of the global wave number 
for the broader class  
of triblock copolymers with $M>6$ segments per block, 
whereas the literature on copolymer mixtures seems to report only the behavior \ref{lifshitz} (see, e.g., \cite{bros_fred90,janert-schick97_2}), associated with a Lifshitz point \cite{hornreich75}.

Since we need to address sequence distributions 
different from the $\lambda$-defined one in the next section, 
we calculate the second-order structure factors from eq.~\xref{S-p-nu} 
for each triblock sequence (class) defined in eq.~\xref{species-def}: 
\begin{subequations}\label{S_nu_M}
\ba
S_{1}(k^2)
&=&D (3M, k^2) \label{S_1_M}\\
&=&
\frac{3M(1+\e^{-k^2})}{1-\e^{-k^2}}-\frac{2\e^{-k^2}(1-\e^{-3Mk^2})}{(1-\e^{-k^2})^2}, 
\nonumber\\
S_{2}(k^2) &=&  - D (3M, k^2)\\
&& \mbox{} + 2 \left( D (2M, k^2) + D (M, k^2) \right), \nonumber\\
S_{3}(k^2) &=&  D (3M, k^2)\nonumber\\ 
&&\mbox{} - 4 D (2M, k^2) + 8 D (M, k^2).
\ea
\end{subequations}
While the maximum of $S_{1}(k^2)$ is located at $k=0$, 
the maximum positions of $S_{2}(k^2)$ and $S_{3}(k^2)$ at $k>0$ 
are due to the finite type-position correlation length within a chain of the
respective sequence. 

The continuous-chain version of the homopolymer structure factor 
eq.\ \xref{S_1_M} is
\be
\label{s_nu_M}
s_{1}(k^2) \mathrel{\mathop:}= \lim_{M\to\infty}
S_{1} \left( k^2 /M \right)/M^2 = g_{\text{D}} (3, k^2),
\ee
again with $k^2\mathrel{\mathop:} = R_{\text{block}}^2
\tilde k^2/(2d)$; similar expressions hold for $s_{2}(k^2)$ and
$s_{3}(k^2)$. In the following, $S_{\nu}(k^2)$ or $S\left( k^2 \right)$ refer
to the discrete structure factors, and the number of segments, $M$, 
is usually not listed as an argument separately.
The continuous versions are denoted with $s_{\nu}(k^2)$,
$s\left( k^2 \right)$ etc. 
Since the number of sequences grows exponentially with $Q$, 
the explicit calculation of sequence-specific structure factors is practically limited to a comparatively
small number of different sequence classes, i.e., to a small number $Q$ of
blocks per chain.

\subsection{\label{sec-lam-phase}Lamellar phase separation}

In order to derive the free energy due to microphase separation, 
we insert for our order-parameter field $\hat\tau$ the simplest single-harmonic ansatz \cite{leibler80}:  
lamellae with wave vector $\bm k_0$, $k_0:=|\bm k_0|>0$ 
and an amplitude $\hat\tau_{\bm k_0}$
\be \hat\tau_{\bm k}=
\hat\tau_{\bm k_0}\,\left(\delta_{\bm k,\bm k_0}+\delta_{\bm k,-\bm k_0}\right) 
\ee
More than one single wave vector is not considered here, since the
instabilities of the disordered state of symmetric copolymers 
are known to be toward homogeneous or lamellar phases. 
In the latter case, we additionally assume that $A$-$B$ separation is weak. 

\subsubsection{Simplified lamellar free energy\label{subsec-lam-simple}}

Insertion of the above ansatz into the simplified functional eq.~\xref{f_4thk0} 
yields the free energy of a lamellar phase,
\begin{align}
\label{fmps1}
\lefteqn{ f_0(k_0, \hat\tau_{\bm
  k_0})}\\
&=\frac{\chi}{2N^2L}\left(
1 - \chi\frac{S\left(k_0^2\right)}{2L}
    \right)
    \hat\tau_{\bm k_0}^2 
    + \frac{\chi^4}{64N^4} 
    \left(m_2^2 + m_4\right) \hat\tau_{\bm k_0}^4,
\nonumber 
\end{align}
which is valid only for
incompatibilities $L\chi$ exceeding
\be 
L\chi_{\text{m}}(k_0) 
= \frac{2L^2}{
S\left(k_0^2\right)}, 
\label{chi-m-nat}
\ee
the onset incompatibility. 
(As usual, we shall use $L\chi$ instead of $\chi$ as one parameter 
of the phase diagrams, due to the scaling of the onset incompatibilities with $L$.)

Minimization of the function eq.~\xref{fmps1} 
with respect to the order-parameter amplitude gives
\be 
\hat\tau_{\bm k_0, \text{m}}^2 = 
\frac{16 N^2 \left( \displaystyle \frac{S\left( k_0^2 \right)}{2L} - \frac{1}{\chi} \right)}{
L\chi^2\left( m_2^2 + m_4 \right)}
\label{sigmafmps1}\ee
Variation with respect to the wave number of the instability 
shows that the optimal $k_0$ is the maximum position of $S(k^2)$.
With the single-harmonic approximation of the profile,
the lamellar free energy 
at $L\chi \geq L\chi_{\text{m}}(k_0)$ is
\begin{align} 
f_{\text{m}} = - \frac{\left({\displaystyle \frac{S(k_0^2)}{L^2} - \frac{2}{L \chi} }\right)^2}{m_2^2 + m_4}, 
\quad & k_0 \mathrel{\mathop:}= \underset{k > 0}{\mbox{argmax}}\, S(k^2).
\label{fmps1_min}
\end{align}
The first two phase diagrams in sec.~\ref{chap-phase-diag} are based on
this simplified version of the lamellar free energy. 

\subsubsection{Lamellar free energy with restored wave-number dependence of fourth order coefficients\label{sec-landau-k4}}

Restoring the $k$-dependence of the fourth-order terms of 
eq.~\xref{f_4th}, 
and optimizing the amplitude at a given wave number $k_{\rm m}$, 
we arrive at the free-energy function
\begin{align}
\label{fmps_compl}
\lefteqn{ f_{\text{m}, k_{\rm m}} = } \\
&
\frac{{\displaystyle
- L^4 \left( \frac{S(k_m^2)}{L^2} - \frac{2}{L \chi} \right)^2}}{
{\displaystyle
\frac{ \left( S^{(\alpha)} \left(\bm k_{\rm m}, \bm k_{\rm m} \right) \right)^2 }{D \left(L, 4k_{\rm m}^2 \right)} - S^{(\beta)}\left(\bm k_{\rm m}, \bm k_{\rm m}, -\bm k_{\rm m}\right) + S^{(\gamma)}\left( k_{\rm m}^2, k_{\rm m}^2\right)}},
\nonumber\end{align}
given $\chi > 2L/S\left(k_{\rm m}^2\right)$.
Now, minimization with respect to $k_{\rm m}$ results in a wave number
that additionally depends on the incompatibility, $k_0 = k_{0}(\chi)$.

\subsection{\label{sec-hom-phases}Macroscopic phase separation}
\subsubsection{Coexistence of two homogeneous phases}

Macroscopic phase separation can be assessed with 
a real-space version of the free-energy functional 
eq.~\xref{f_4thk0}. Accounting for the symmetry, the appropriate ansatz is for two 
phases with uniform fields $\hat\tau$ of opposite signs 
in equally sized regions $V_{\text{h}, 1}$ and $V_{\text{h}, 2}$ of the system:
\be \hat\tau(\bm x) = \left\{ \begin{array}{ll} 
\hat\tau_{\text{h}}, & \bm x\in V_{\text{h}, 1}\\
- \hat\tau_{\text{h}}, & \bm x\in V_{\text{h}, 2}
\end{array}
\right\}, |V_{\text h, 1}| = |V_{\text h, 2}| = \frac{V}{2}
\label{2phase-ans}
\ee
With this ansatz, 
the free energy of Landau form becomes
\be\label{fh_landau}
f_{\text{h}, 0} =
\frac{L \chi}{4\varrho_0^2} \left(1-\frac{\chi S(0)}{2 L}\right)\hat\tau_{\text{h}}^2
+\frac{(L\chi)^4 m_4}{192\varrho_0^4}\hat\tau_{\text h}^4,
\ee
which provides a good description of macroscopic phase separation for
small values $\hat\tau_{\text h}$
close to the continuous transition from the disordered state. 

However, the transition we aim at, 
from the macroscopically phase-separated to a three-phase state, 
may occur 
at a value $L\chi$ considerably larger than the onset incompatibility 
of macroscopic phase separation;
see fig.~\ref{fig-chi2m-chih-M3}. 
Thus instead of the free energy eq.~\xref{fh_landau} 
that relies on an expansion in $\hat\tau_{\text h}$,
we prefer and are able to derive a closed expression (cf.~appendix \ref{app_crushed-polymers})
by ignoring the copolymers' internal structure, consistent with uniform mean fields.
For random triblock copolymers, the free energy is
\begin{align}
f_{\text{h}}= 
&\frac{L \chi \hat\tau_{\text{h}}^2}{4 \varrho_0^2}
- \left(1- p_1 \right)
\ln\cosh \frac{L\chi\hat\tau_{\text{h}}}{6\varrho_0}
- p_1
\ln\cosh \frac{L\chi\hat\tau_{\text{h}}}{2\varrho_0},
\nonumber\\
& \mbox{ provided }
L\chi > L\chi_{\text{h}} \mathrel{\mathop:}= \frac{2}{ m_{2} } = 
\frac{18}{ 1 + 8 p_1 }, 
\label{fh_multiph}
\end{align}
with the homopolymer concentration, $p_1 
= (1 - p_2 - p_3)$
(the indices $2$ and $3$ refer to the sequence classification 
eq.~\xref{species-def} needed in the description of a lamellar phase). 
Here, the amplitude $\hat\tau_{\text{h}}$ is determined by
\be
\label{sigma_h}
\frac{\hat\tau_{\text{h}}}{\varrho_0} = 
\frac{1 - p_1}{3}\, \tanh \frac{L\chi\hat\tau_{\text{h}}}{6\varrho_0}
+ p_1 \tanh \frac{L\chi\hat\tau_{\text{h}}}{2\varrho_0}.
\ee

\subsubsection{\label{4phases}Homogeneous multi-phase coexistence}
Within multi-component theory, the two homogeneous, $A$- and $B$-rich phases in a random triblock melt 
(formed at $L \chi_{\text{h}} = 6$ for $\lambda = 0$) 
are followed by
\emph{four\/} homogeneous phases at higher incompatibilities (e.g., $L \chi_{\text{h, 4}} \approx 16$
for $\lambda = 0$). More than four phases are impossible within this theory, since for $Q=3$ 
there are only four different chain compositions (A contents). 
The fact that no triblock sequence is symmetric in A/B content might explain why, starting from the $A$- and the $B$-rich phase, a third, homogeneous phase balanced in A/B does not become stable.

\section{\label{chap-f-frac}Fractionated three-phase coexistence}

In the following, we show that both a global macrosopic and a global
lamellar phase separation become unstable toward three-phase coexistence due to fractionation. 
In the former case, mainly alternating sequences are expelled from the
macroscopically phase-separated state (cloud) to allow for a third lamellar
shadow phase, whereas in the latter case mainly homopolymers are expelled
from the lamellar state (cloud) to allow for two additional homogeneous shadow phases.

\subsection{\label{sec-f-hom}Fractionation from two macroscopic phases}

Here, we start at block-type correlations $\lambda > \lambda_c$ 
and incompatibilities $L\chi>L\chi_{\text{h}}$, 
i.e., from a 
global, macroscopically phase-separated state comprising
two homogeneous $A$-rich, respectively $B$-rich, phases. 
At further increase of $L\chi$, 
a third, lamellar phase with zero average $A$ excess 
will be created by fractionation: 
Predominantly alternating sequences
($\nu = 2,3$) with few {homo}\-polymers 
will remix in a volume fraction $v^{(2)}\mathrel{\mathop:}= V^{(2)}/V$ of the system.
For our symmetric distributions, the two homogeneous phases
coincide in the volume fractions, in the field values up to the sign,
in the sequence (class) concentrations defined in eq.~\xref{species-def}, 
and thus in the free-energy densities. 
Hence we can treat them as one effective state, 
and study their \emph{joint\/} sequence exchange with a lamellar phase.

The first term of the free energy relative to the homogeneous, two-phase state,
is written as a weighted sum of the free-energy densities $f_{\text{m}}^{(2)}$ 
of the conjectured lamellar phase with volume fraction $v^{(2)}$, 
and $f_{\text{h}}^{(1)}$ of the two homogeneous, homopolymer-rich phases
with joint volume fraction $v^{(1)} = 1- v^{(2)}$: 
\be\label{f_sum} 
f_{\text{sum}} \mathrel{\mathop:}=
v^{(2)}f_{\text{m}}^{(2)}\left(\left\{n_{\nu}^{(2)}\right\}\right)
 +  (1-v^{(2)})f_{\text{h}}^{(1)}\left(\left\{n_{\nu}^{(1)}\right\}\right). 
\ee
Here, the sequence concentrations in state (phase) $P$
are denoted as $n_{\nu}^{(P)},\, \nu=1,2,3,\, P=1,2$. 
The free-energy densities of \emph{global\/} ordered states alone 
(for which combinatorial terms due to the sequence distribution cancel)
cannot completely describe the coexistence of different states 
that interact via sequence exchange.
Hence there are 
additional entropic coupling terms:
First, confinement of the chains to the volume \emph{fractions\/} 
of phase-separated subsystems gives rise to a loss
\be\label{f_vol} 
\Delta f_{\text{vol.\ red.}} \mathrel{\mathop:}= 
 -v^{(2)}\ln v^{(2)} -(1-v^{(2)})\ln(1-v^{(2)})
\ee
of translational entropy compared to the global state.
Second, the sequence-selective exchange between the two
phases effects a combinatorial gain $\Delta f_{\text{comb.}}$ 
due to the possibilities 
to choose chains of each sequence in one subsystem out of the
total, $\lambda$-defined number $N p_{\nu}(\lambda)$ 
(the factorials are approximated by Stirling's formula):
\ba\label{f_comb}
\lefteqn{
\Delta f_{\text{comb.}}\mathrel{\mathop:} =
\sum_{\nu = 1}^{3}
\Biggl\{
v^{(2)} n_{\nu}^{(2)} 
\ln\left[\frac{v^{(2)} n_{\nu}^{(2)}}{p_{\nu}(\lambda)}\right]
}\\ && 
\mbox{}  + \left( 1-v^{(2)} \right) n_{\nu}^{(1)} 
\ln\left[\frac{\left( 1-v^{(2)} \right) n_{\nu}^{(1)}}{p_{\nu}(\lambda)}\right]
\Biggr\}. 
\nonumber
\ea
With the above contributions, the free energy 
of the fractionated phase coexistence
is
\be\label{f-frac} 
f_{\text{frac}}
=f_{\text{sum}}  + \Delta f_{\text{vol.\ red.}} + \Delta f_{\text{comb.}}.
\ee

Incompressibility and the global $\lambda$-defined sequence distribution
reduce the number of variables, 
given by the volume fraction $v^{(2)}$ of the lamellar phase and 
the concentrations $n_{\nu}^{(P)},\, \nu=1,2,3,\, P=1,2$:
The homopolymer concentrations $n_{1}^{(k)}$ 
can be eliminated by the 
constraints 
\be\label{n-conserv} 
\sum_{\nu=1}^3 n_{\nu}^{(P)} = 1 \mbox{ for each phase } P=1,2. 
\ee
Likewise, the concentrations $n_{\nu}^{(1)}$ in the homogeneous phases 
can be explicitly expressed in terms of the volume fraction and the
concentrations in the lamellar phase via the constraint
of global $\lambda$-defined concentrations,
\ba
v^{(2)}n_{\nu}^{(2)}+(1-v^{(2)})n_{\nu}^{(1)} = p_{\nu}(\lambda),\, \nu=2,3.
\label{constraint-p-lambda}
\ea
Thus left with three independent variables, 
we choose them as $v^{(2)}$, $n_2^{(2)}$, and $n_3^{(2)}$ for the purpose of studying fractionation starting from two homogeneous phases.

Obviously, at a given block-type correlation $\lambda$ 
and a given incompatibility $L\chi$, the fractionation ansatz eq.\
\xref{f-frac} is 
reasonable only for values of the variables $v^{(2)}$, $n_2^{(2)}$, $n_3^{(2)}$ for which
$f_{\text{frac}}$ reaches lower values 
than the free-energy density $f_{\text{h}}$ 
of the global 
state: 
\be \label{Df-frac}
\Delta f_{\text{frac}}\left( v^{(2)}, n_2^{(2)}, n_3^{(2)} \right) 
\mathrel{\mathop:}= 
f_{\text{frac}} - f_{\text{h}}\overset{!}{\leq} 0.
\ee
For each set $(\lambda, \chi)$, the free-energy change
$\Delta f_{\text{frac}}$ has to be minimized with respect to
$v^{(2)}$, $n_2^{(2)}$, $n_3^{(2)}$ within the region limited by eq.~\xref{Df-frac}. 
To avoid overloading the presentation, 
the functional dependence on $\lambda$ and $L\chi$  is suppressed in $f_{\text{frac}}$, 
as well as in the free-energy densities of the global homogeneous and lamellar phases.

To obtain the free energy of the lamellar state with fractionation, 
we compute the structure factor eq.~\xref{S-p-nu} and
the moments eq.~\xref{nth-mom} with modified sequence concentrations
$p_2 \to n_2^{(2)}$ and $p_3\to n_3^{(2)}$ ($p_1=1-p_2-p_3$), 
which are then the explicit arguments of $f_{\text{m}}^{(2)}$.
Similarly, the free-energy density of the macroscopically 
phase-separated state with fractionation is computed with
modified concentrations $n_2^{(1)}$, $n_3^{(1)}$, such that, 
via eq.~\xref{constraint-p-lambda},
$f_h^{(1)} 
$ becomes a function of
$v^{(2)}$, $n_2^{(2)}$, and $n_3^{(2)}$.

\subsection{\label{sec-f-lam}Fractionation from a global lamellar phase}

The assumed boundary curve between the three-phase coexistence region and
one lamellar phase comprising the total system, 
is in our approach restricted to the region $\lambda < \lambda_c$ of the $\lambda$-$\chi$ space.
To access this region, 
a fractionation ansatz has to start from lamellae in the $\lambda$-distribution,
which tend to expel homopolymers on increasing $\chi$, 
a mechanism which will give rise to homogeneous $A$- and $B$-rich shadow phases.
The fractionation free energy is formulated in analogy to eq.~\xref{f-frac} 
in terms of the free-energy densities of 
one effective homogeneous shadow phase, in a volume fraction $v^{(1)}$,
and a lamellar cloud phase, in a volume fraction $1 - v^{(1)}$. 
For $v^{(1)} > 0$, both states attain sequence concentrations deviating from 
the $\lambda$-defined ones. 
The first part of the fractionation free energy corresponding to eq.\ \xref{f_sum} is
\begin{align}
\lefteqn{f_{\text{sum}} =} \\
& v^{(1)} f_{\text{h}}^{(1)} \left( n_2^{(1)}, n_3^{(1)} \right)
+ (1- v^{(1)}) f_{\text{m}}^{(2)} \left( v^{(1)}, n_2^{(1)}, n_3^{(1)} \right).
\nonumber\end{align}
Again, the constraints eqs.~\xref{n-conserv} and \xref{constraint-p-lambda} 
of incompressibility and fixed global
sequence distribution reduce the number of independent variables to 3;
in this case, they are chosen as $ v^{(1)}$, $n_2^{(1)}$, $n_3^{(1)}$. 
The entropic terms due to a loss of translational
entropy 
and due to combinatorial gains 
by three-phase coexistence are constructed in complete analogy to
eqs.~\xref{f_vol} and \xref{f_comb}.

\subsection{\label{sec-eq-cond}Fractionated three-phase equilibrium conditions}

We minimize the fractionation free energy 
presented in the last subsections with respect to 
the volume fraction and sequence distribution of the emerging shadow phase(s).
Insertion of the free-energy densities of the different states with fractionation 
into eq.~\xref{f-frac} and subsequent differentiation of $f_{\text{frac}}$
with respect to the variables $v^{(2)}$, $n_2^{(2)}$, $n_3^{(2)}$ or $v^{(1)}$,
$n_2^{(1)}$, $n_3^{(1)}$ 
give equation systems
\be\label{minimization_f}
\bm 0 = \left( 
\frac{\partial f_{\text{frac}}}{\partial v^{(P)}}, 
\frac{\partial f_{\text{frac}}}{\partial n_2^{(P)}}, 
\frac{\partial f_{\text{frac}}}{\partial n_3^{(P)}}
\right),\quad P=1,2
\ee
exemplified in appendix \ref{app-min-cond}.
Solutions are obtained numerically with a Newton-type
procedure (cf.~appendix~\ref{numerical}).

\subsection{\label{subsec-minv01}Three-phase transition lines}

Upon gradually decreasing or increasing the incompatibility 
$L\chi$ at fixed $\lambda$ in the three-phase state,
boundaries of the three-phase region,
$\chi^{(1)}$ at $\lambda > \lambda_c$, and
$\chi^{(2)}$ (at $\lambda < \lambda_c$ with our simplified lamellar free energy), 
are indicated by a zero of the free energy eq.~\xref{f-frac} due to fractionation:
Either the minority phase's volume fraction tends to zero (characteristic of a shadow), 
its sequence concentrations approach the $\lambda$-defined ones, 
or its order-parameter amplitude tends to zero.
Analysis of eq.~\xref{minimization_f} shows that in our system
the first alternative is realized, which simplifies the set of equations for the transition lines.
In the case of 
sec.~\ref{sec-f-hom}, an expansion of the entropic contributions eqs.~\xref{f_vol}, \xref{f_comb} to 
$f_{\text{frac}}$ 
in the volume fraction $v^{(2)}$ of the lamellar shadow phase yields
\ba\label{exp_f_v2}
\lefteqn{\Delta f_{\text{vol.\ red.}}+\Delta f_{\text{comb.}}}\\
&=&
\sum_{\nu=1}^{3}n_{\nu}^{(2)}\ln\left( \frac{n_{\nu}^{(2)}}{p_{\nu}(\lambda)} \right) 
\,v^{(2)}
\nonumber\\
&& \mbox{} + \frac{1}{2}\,\sum_{\nu=1}^{3}n_{\nu}^{(2)}\left(
\frac{n_{\nu}^{(2)}}{p_{\nu}(\lambda)}-1\right) 
\,\left(v^{(2)}\right)^2
+  {\cal{O}} \left( \left( v^{(2)} \right)^3 \right).
\nonumber\ea
Similarly, 
one can expand the deviations from $\lambda$-defined concentrations in the 
two-phase cloud state:
\begin{align} 
\label{exp_n_v2}
n_{\nu}^{(1)}  - p_{\nu}(\lambda) 
=& \left( p_{\nu}(\lambda)-n_{\nu}^{(2)} \right)\, v^{(2)}\\
&  \mbox{} + \left( p_{\nu}(\lambda)-n_{\nu}^{(2)} \right)\, \left( v^{(2)} \right)^2 
+ {\cal{O}} \left( \left( v^{(2)} \right)^3 \right) 
\nonumber
\end{align}
Hence the lowest-order term of $f_{\text{frac}}$ is linear in $v^{(2)}$,
\begin{align}
\lefteqn{
\Delta f_{\text{frac}}(v^{(2)}, n_2^{(2)}, n_3^{(2)}) }\label{d-ffrac-expansion}\\
&= a(n_2^{(2)},
n_3^{(2)}) v^{(2)}+{\cal{O}} \left( ( v^{(2)} )^2 \right),
\nonumber\end{align}
and the coefficient $a$ must be minimized
in order to determine $\chi^{(1)}$ 
and the sequence distribution of the shadow phase.
The phase transition line from the global lamellar to the fractionated three-phase state 
can be treated in complete analogy by taking the limit $v^{(1)}\to 0$.  

\section{\label{chap-phase-diag}Phase diagrams}

In the following, we present the lines of macroscopic and lamellar phase separation 
and the boundary lines of three-phase coexistence
obtained from the minimization of the fractionation free energy.
The critical line of macroscopic phase separations of the disordered melt is well known already from approaches based on 
the multi-component picture \cite{bauer85, nesarikar}.
Also the discussion of the pure microphase separation transition within mean-field theory can be found elsewhere \cite{leibler80,fred92,erukhimov96}. Our focus here is on the coexistence of homogeneous and lamellar phases with fractionated sequence distributions.
The point $(\lambda_c, L\chi_c)$ 
where the transition curves from the disordered toward macroscopic, respectively 
lamellar, phase separation meet
will be mostly referred to as a multicritical point without further classification. 
Partitioning of sequences will be shown via distribution diagrams in sec.~\ref{sec-fract-distr}, 
in comparison with SCFT calculations.

\subsection{\label{sec-discreteM3}Triblocks with small $M$ 
}

\begin{figure}[h!]
\includegraphics[width=\columnwidth]{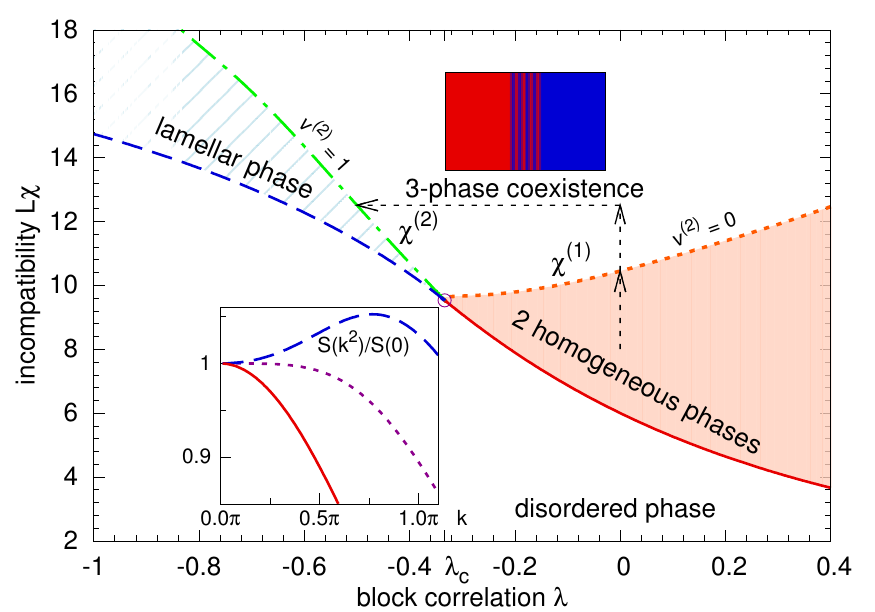}
\caption{(Color online) Phase diagram for triblock copolymers with $M=3$. 
Solid (red) line marks macroscopic (two homogeneous, $A$- and $B$-rich phases),  
dashed (blue) line marks lamellar phase separation (order-disorder-transition ODT)
of the disordered state. 
Dotted (orange) line: onset of three-phase coexistence, 
at which the two homogeneous 
phases are the cloud and
a fractionated lamellar phase shadow emerges; 
dot-dashed (green) line is the lamellar cloud boundary.
A circle 
marks the multicritical point.
Bottom inset: second-order structure factor in the global $\lambda$-distribution, 
at $\lambda = 0$ (solid), 
at the critical correlation $\lambda_c = -\nicefrac{1}{3}$ (dotted), 
and at $\lambda = -0.5$ (dashed).
In this and the following plots of this part, the length scale is $\left\langle\left(\bm n\cdot\bm R_e\right)^2\right\rangle^{1/2} =R_e/\sqrt{d}$.
Top inset: sketch of three-phase coexistence.
\label{fig-chi2m-chih-M3}}
\end{figure}

To exemplify the phase behavior of triblocks with $M < 7$ segments per block, 
we discuss the results for $M=3$ shown in fig.\ \ref{fig-chi2m-chih-M3}.
To explore the emergence and growth of the various phases, we follow the path indicated by arrows in the plot, starting at a block correlation $\lambda > \lambda_c = - \nicefrac{1}{3}$:
The first instability of the disordered melt is toward homogeneous phase separation, indicated by
the peak at zero wave number
of the global, second-order structure factor  (cf.\ the solid curve in the bottom inset). 
Upon increasing incompatibility $L\chi$ (bottom vertical arrow), the dotted line ($\chi^{(1)}$) marks the onset of three-phase coexistence via a fractionated lamellar shadow phase with volume fraction $v^{(2)} = 0$. 
(A fractionated lamellar shadow was already predicted by Monte Carlo simulations \cite{mmueller04}.) 
This fractionated phase sets in with finite amplitude, and with \emph{finite wave number\/}, 
since its copolymer-enriched sequence distribution (see fig.~\ref{fig-conc_triangle} below)
causes the structure factor to be different from the global one. 
On further increase of the incompatibility (along the top vertical arrow), the lamellar volume fraction grows. 
Now, keeping $L\chi$ constant, and proceeding toward smaller values of $\lambda$ 
(following the horizontal arrow), the volume fraction of the lamellae increases further. Finally, at some $\lambda < \lambda_c$, one reaches the end of the three-phase coexistence (indicated by the dot-dashed line), and lamellae take over to be the cloud phase with volume fraction $v^{(2)} = 1$.  
Consistently, starting at $\lambda < \lambda_c$ and small incompatibilities, the disordered melt undergoes lamellar phase separation (at the incompatibilities on the dashed line) due to the peak of the $\lambda$-defined structure factor at a finite wave number 
(cf.\ eq.~\xref{fmps1_min} and the dashed curve in the bottom inset).
With our simplified free energy eq.~\xref{f_4thk0}, via which the instability toward a global lamellar phase rests solely on the $k$-dependence of this second-order structure factor, 
the lamellar cloud boundary of three-phase coexistence is always located in the half-plane $\lambda \leq \lambda_c$. 
Upon crossing the dot-dashed boundary line from this side, two additional homogeneous phases with homopolymer-enriched sequence distributions 
appear as shadows. 

As visible in fig.~\ref{fig-chi2m-chih-M3}, three-phase coexistence prevails in a large parameter region. However, since our lamellar free energy is limited to small order-parameter amplitudes, the results may be unreliable 
at very large values of the incompatibility. 
An alternative scenario would be 
global lamellar phase separation at higher $L\chi$ (see fig.~\ref{ph-diag-scft} below).

At the critical correlation $\lambda_c$, the maximum at $k_0=0$ of the global structure factor broadens (see dotted curve in the bottom inset in fig.\ \ref{fig-chi2m-chih-M3}), announcing the continuous growth of the optimal wave number from zero when lowering $\lambda$. 
Qualitatively, we observe this transition from global macroscopic to global lamellar phase separation for all random triblock copolymers with $M<7$ (cf.\ the case discussed before eq.~\xref{eq-lambda_c}), while the exact position of the Lifshitz point $(\lambda_c, L\chi_c)$ depends on $M$. This point of diverging lamellar wavelength limits the three-phase region toward low incompatibilities.

\begin{figure}[h!]
\includegraphics[width=8cm]{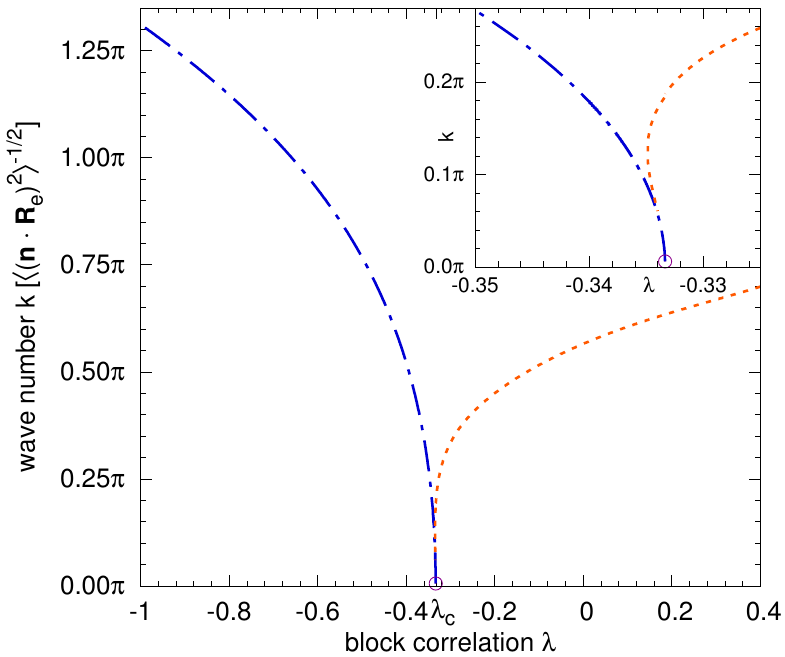}
\caption{(Color online) Lamellar wave numbers 
for triblocks with $M=3$. 
Dot-dashed (blue) line: global lamellar phase at $\lambda < \lambda_c$, 
between the order-disorder transition 
and the onset of three-phase coexistence with $v^{(2)} = 1$
(hatched region of fig.~\ref{fig-chi2m-chih-M3}). 
Dotted (orange) line: fractionated lamellar shadow ($v^{(2)} = 0$).
Inset: enlarged around the multicritical point. 
\label{fig-kopt_M3}}
\end{figure}

The lamellar wave numbers on the boundary lines of fractionated three-phase coexistence 
as a function of $\lambda$ are displayed in fig.\ \ref{fig-kopt_M3}. 
Note that the simplified free energy for microphases (see eq.~\xref{fmps1_min}), 
predicts that at a given $\lambda$,  the wave number of global lamellar phase separation 
(hatched region in fig.~\ref{fig-chi2m-chih-M3}) does not change with increasing $L\chi$. 
The lamellar wave number can be shifted only due to fractionation, 
i.e., by an increased content of alternating sequences. 
We find that, on increasing $L\chi$ in the three-phase region, the fractionation and thereby the wave number in the lamellae increase, i.e., the lamellar spacing decreases. 
This is in agreement with findings for global microphase separation in random copolymers within mean-field theory \cite{fred-miln91}.

\begin{figure}[h!]
\includegraphics[width=7.5cm]{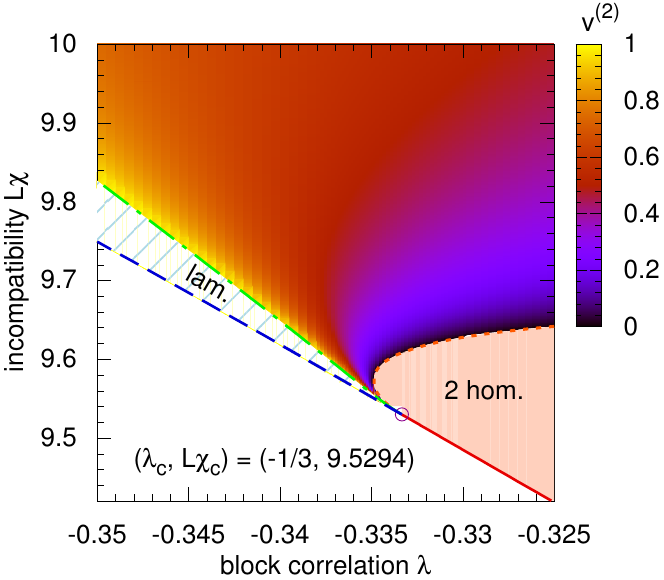}
\caption{(Color online) Volume fraction of lamellar phase 
and three-phase coexistence lines around the multicritical point $(\lambda_c, L\chi_c)$ for triblocks with $M=3$. Line styles as in fig.~\ref{fig-chi2m-chih-M3}.
  \label{fig-v2-Nchi-M3}}
\end{figure}

The wave number of fractionated lamellae vanishes at the Lifshitz point 
$\left( \lambda_c,\chi_c\right)$, as does the wave number of global lamellar phase separation.
The inset in fig.~\ref{fig-kopt_M3} shows the behavior of the fractionated wave number in the vicinity of the Lifshitz point. For $\lambda \lesssim \lambda_c$, the three-phase region can be entered at two different incompatibilities, with different wave numbers of the lamellar shadow.
A closer look is cast onto 
this remarkable feature of the phase diagram 
in the 
detail of the boundary lines and a map of the lamellar phase's volume fraction around the Lifshitz point in fig.~\ref{fig-v2-Nchi-M3}.
The line of fractionated lamellar shadows 
displays a \textit{reentrant\/} behavior, 
especially it does not reach the Lifshitz point for $\lambda \searrow \lambda_c$, 
but via a spiraling path invading the region $\lambda < \lambda_c$. 
Except for a very small region of the parameter space, 
fractionation suppresses global lamellae with diverging wavelength 
in the vicinity of the Lifshitz point, in favor of, first, macroscopic phases and, at higher incompatibilities, fractionated lamellae with finite wavelength.

\begin{figure}[h!]
\includegraphics[width=8cm]{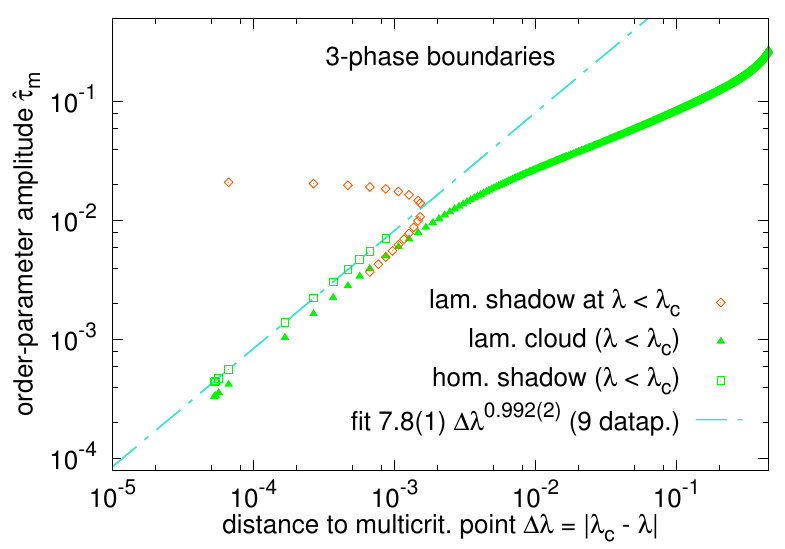}
\caption{(Color online) Lamellar order-parameter amplitude 
along the boundaries of three-phase coexistence for $M=3$.
\label{fig-ampl-lambda_c_M3}}
\end{figure}
The scaling of the order-parameter amplitude on approach to the multicritical point 
along the transition lines to three-phase coexistence ($\lambda \nearrow \lambda_c$) is shown in
fig.~\ref{fig-ampl-lambda_c_M3}.  
The amplitudes of fractionated lamellar shadows (on the 
dotted line in fig.~\ref{fig-v2-Nchi-M3}, in the range $\lambda < \lambda_c$) are marked by open diamonds, 
those of global lamellar (cloud) phases (on the the dot-dashed line in fig.~\ref{fig-v2-Nchi-M3}) 
by solid triangles, those of the coexisting homogeneous shadows by open squares. 
According to the fit performed to the latter case, the amplitudes
vanish linearly in $\Delta\lambda  \mathrel{\mathop:} = |\lambda - \lambda_c|$ at the Lifshitz point
(the same exponent is found for the lamellar cloud amplitude).

In order to analytically extract the exponent of the 
order-parameter amplitude in the vicinity of the multicritical point, 
we solve the equation of the lamellar cloud line $v^{(1)} = 0$  (see sec~\ref{subsec-minv01})
for the deviations of the sequence concentrations in the fractionated macroscopic shadow phases from the global ones, 
$\Delta n_{\nu} \mathrel{\mathop:} = n_{\nu}^{(1)} - p_{\nu}(\lambda)$, with a power series ansatz 
\be 
\Delta n_{\nu} (\Delta\lambda) =\sum_{j} c_{\nu j} \left( \Delta \lambda\right)^j.
\label{deltan-pow}\ee
The series' coefficients of the equation system in $\Delta\lambda$ can be calculated for $M\in[3,\ldots,6]$, cases in which the wave number $k_0$ of global lamellae vanishes $\propto \left( \Delta\lambda\right)^{1/2}$ at $\left(\lambda_c,\chi_c\right)$ (as expected for a Lifshitz point \cite{fred92}). For $M=3$, consistent expansion up to $\left( \Delta\lambda \right)^4$ yields, 
along the lamellar cloud boundary line,
\be \Delta n_{\nu} = -\frac{144\sqrt{6}}{55}\left( \Delta\lambda \right)^2 +  {\cal{O}}\bigl( (\Delta \lambda)^3 \bigr),\quad \nu = 2, 3.\ee
When inserting these dependencies 
into expansions of the optimal wave number, the structure factor etc.\ (cf.\ eq.~\xref{sigmafmps1}),
we indeed find the critical exponent $1$ 
for the amplitude $\sigma_{\text m}$ 
in the lamellar cloud phase, 
 \be \label{crit-exp1analyt}
 \hat\tau_{\text m} \propto \Delta\lambda, \quad\lambda\nearrow \lambda_c. 
 \ee
Moreover, the slopes of the transition lines $L\chi(\lambda)$ from the disordered to the global lamellar state ($\chi_{\text m}(\lambda)$) and from global lamellae to fractionated three-phase coexistence ($\chi^{(2)}(\lambda)$) 
can be shown to be equal at $(\lambda_c, L\chi_c)$:
\be
\chi^{(2)}(\lambda) - \chi_{\text m}(\lambda)
 \propto \left(\Delta\lambda\right)^2, \quad \lambda \nearrow \lambda_c. 
\ee

\subsection{\label{sec-cont}Continuous triblocks}

\begin{figure}[h!]
\includegraphics[width=\columnwidth]{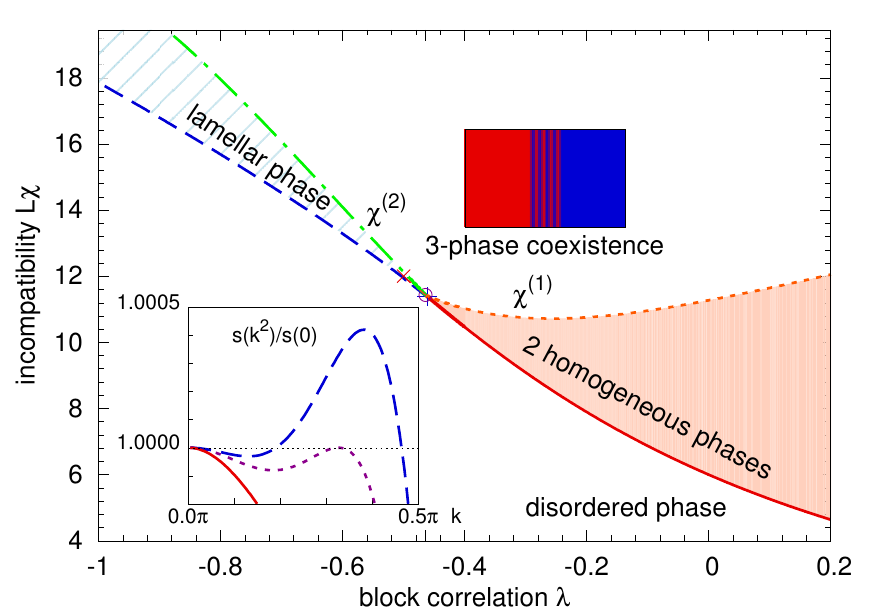}
\caption{(Color online) 
Phase diagram for continuous triblocks. 
Line styles as in fig.~\ref{fig-chi2m-chih-M3}.
Crosses indicate the end points of the lines of metastable, global phase separations, 
macroscopic  for $\lambda < \lambda_c$ ($\color{red}\times$) and lamellar for $\lambda > \lambda_c$ ($\color{blue}+$).
Bottom inset: global second-order structure factor 
at $\lambda = -0.45$ (solid), 
at the critical correlation $\lambda_c = -0.464\,00$ (dotted),
and at $\lambda = -0.47$ (dashed). Length scale as in fig.~\ref{fig-chi2m-chih-M3}.
\label{fig-chi2m-chih-cont}}
\end{figure}

Representative of triblocks with $M \geq 7$ segments per block, 
the phase diagram for continuous random triblock melts is shown 
in fig.\ \ref{fig-chi2m-chih-cont}.
Again, for $\lambda > \lambda_c$, the dotted line marks the emergence of 
a lamellar shadow in addition to the two homogeneous phases. 
The lamellar volume fraction grows with increasing $L\chi$ and with decreasing $\lambda$. 
On the dot-dashed line, the lamellar phase takes over to be the cloud phase and coexists with two fractionated homogeneous shadows.

\begin{figure}[h!]
\includegraphics[width=7.5cm]{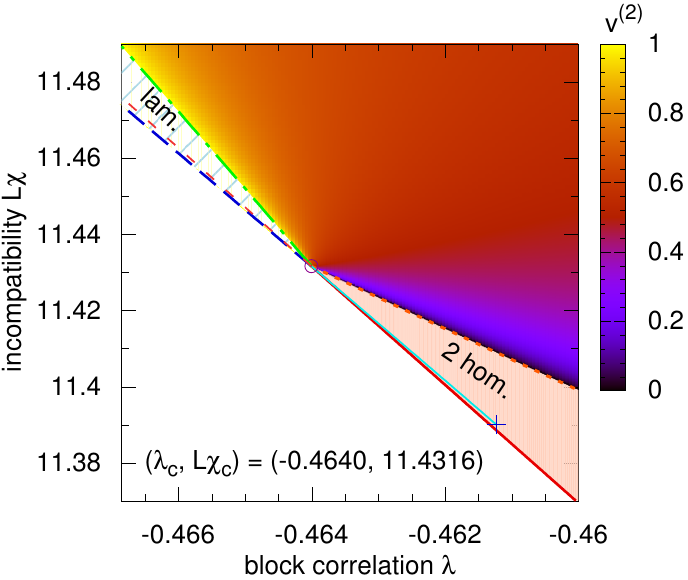}
\caption{(Color online)
Volume fraction  of the lamellar phase around the multicritical point 
$(\lambda_c, L\chi_c) 
$
for continuous triblocks. Boundary lines of three-phase coexistence 
as in fig.~\ref{fig-chi2m-chih-cont}. 
Additional thin lines mark metastable, global phase separations: 
solid (cyan) for $\lambda > \lambda_c$, with end point marked by a cross: 
lamellar phase;
dashed (red) for $\lambda < \lambda_c$:
homogeneous phases.
\label{fig-v2-Nchi-cont}}
\end{figure}

In comparison to the case $M<7$ (see fig.~\ref{fig-chi2m-chih-M3}), 
the three-phase coexistence region seems to be larger. (Still, the predictions are restricted
to incompatibilities that do not exceed considerably those of the order-disorder transition.)
The multicritical point is not only located 
at a smaller critical block correlation $\lambda_c$ and a higher incompatibility, 
but is also qualitatively different:
As discussed in sec.~\ref{sec-S-k-lambda}, the 
wave number of the first global, ordered structure 
(when starting at low incompatibilities in the disordered state) 
is \emph{discontinuous\/} at $\lambda_c$ for $M \geq 7$. 
Thus when reaching $\lambda_c$ from above, 
the morphology of the ordered phase 
changes from two homogeneous phases (zero wave number $k_0=0$) 
to one lamellar phase with finite wave number $k_{0,c}$. 
\begin{figure}[h!]
\includegraphics[width=8cm]{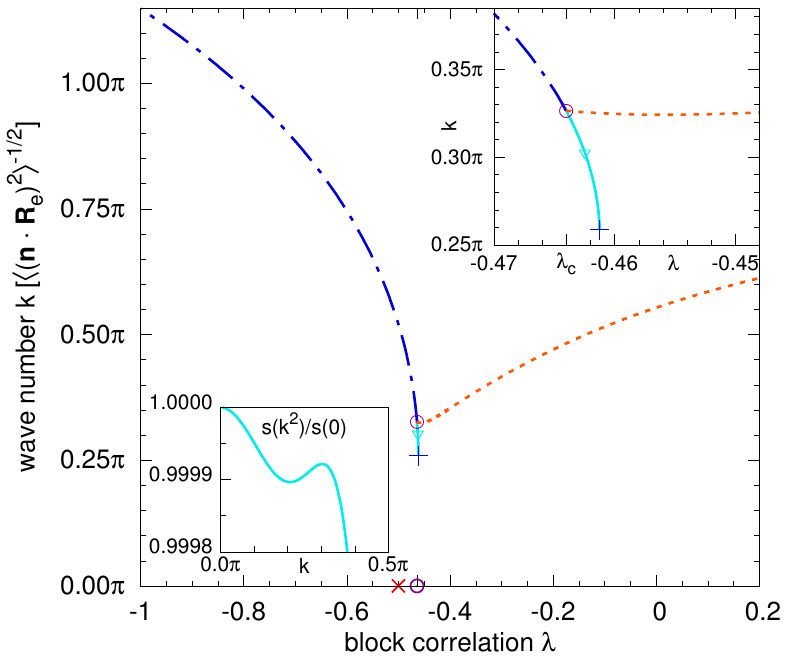}
\caption{(Color online)
Lamellar wave number for continuous triblocks. 
Dot-dashed (blue) line: global lamellar (cloud) phase at $\lambda < \lambda_c$; 
dotted (orange) line: fractionated lamellar shadow ($v^{(2)} = 0$); 
solid (cyan) line: metastable global lamellae due to a second peak 
of the structure factor $s(k^2)$, shown at $\lambda = -0.4625$ (triangle)
in the bottom inset. Circles mark the wave numbers of the coexisting states at the multicritical point,
crosses mark the end points of metastable global lamellar/macroscopic phase separation lines. 
\label{fig-kopt_cont}}
\end{figure}
This feature is revealed in more detail in the plot of lamellar wave numbers 
in fig.~\ref{fig-kopt_cont}.  
At the multicritical point, the lamellar wave numbers in the fractionated state
also tend to the finite value $k_{0,c} 
= 0.326\pi$. 
The wave number in the fractionated lamellar shadow 
attains a slightly smaller, minimal value at a correlation $\lambda > \lambda_c$
(cf.\ the top inset in fig.~\ref{fig-kopt_cont}). 
Due to the \emph{two\/} peaks
of the global structure factor around multicriticality 
(see the inset in fig.~\ref{fig-kopt_cont}),
metastable global lamellae occur in a small range of block correlations,
$-0.461\,23 \geq \lambda > \lambda_c$,
where the free-energy functional's absolute minimum 
indicates global macroscopic phase separation.
Inversely, global macroscopic phase separation persists as a metastable state 
for $-0.5 < \lambda < \lambda_c$ 
(at $\lambda = -0.5$, the curvature of $s(k^2)$ at $k=0$ changes according to eq.\ \xref{eq-lambda_c}).
These metastable transition lines, 
whose end points are hardly resolvable in fig.~\ref{fig-chi2m-chih-cont}, 
are displayed in fig.~\ref{fig-v2-Nchi-cont}, 
together with the actual transition lines and a map of the lamellar volume fraction
around the multicritical point 
(note the zoom to an even smaller region than in fig.~\ref{fig-v2-Nchi-M3}).
On increasing incompatibility from the fractionation onset, 
the lamellar volume fraction grows (for $\lambda > \lambda_c$) or decreases (for $\lambda < \lambda_c$) 
rapidly to level out at a value of about $0.6$.
At multicriticality,
the transition lines from the disordered to the global lamellar state and from global lamellae to fractionated three-phase coexistence differ in their slopes, 
in contrast to the behavior at the Lifshitz point. 

\begin{figure}[h!]
\includegraphics[width=8cm]{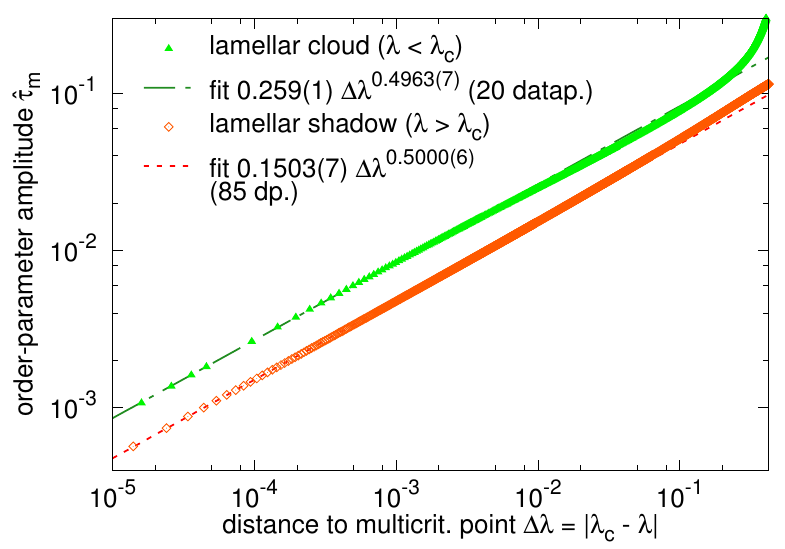}
\caption{(Color online) Scaling of the lamellar order-parameter amplitude along three-phase boundaries
for continuous triblocks. 
\label{fig-ampl-lambda_c_cont}}
\end{figure}
Despite the discontinuity of the wave number at the critical correlation $\lambda_c$ for $M\geq 7$, 
the boundary lines of fractionated three-phase coexistence 
are single-valued around the multicritical point. 
Hence, in this case, we can determine numerically
the 
critical exponent for the decay of the lamellar order-parameter amplitude 
along both boundary lines (see fig.~\ref{fig-ampl-lambda_c_cont}). 
The exponent $0.5$, found along both lines, 
is reminiscent of 
mean-field behavior.
Note that for triblocks with $M<7$, we derived a different critical exponent, viz.\ $1$ (cf.\ eq.~\xref{crit-exp1analyt}).

\subsection{Fractionation with restored wave number dependence\label{sec-frac-compl}}

In this section, we aim at testing the fractionation scenario 
with the complete fourth-order expansion of the  Landau-type free energy for structured phases, 
instead of the simplified version eq.~\xref{f_4thk0}.
To this end, we accounted for the wave number dependence of the fourth-order terms of the functional eq.~\xref{f_4th} in eq.~\xref{fmps_compl} in sec.~\ref{sec-landau-k4}. 
\begin{figure}[h!]
\includegraphics[width=\columnwidth]{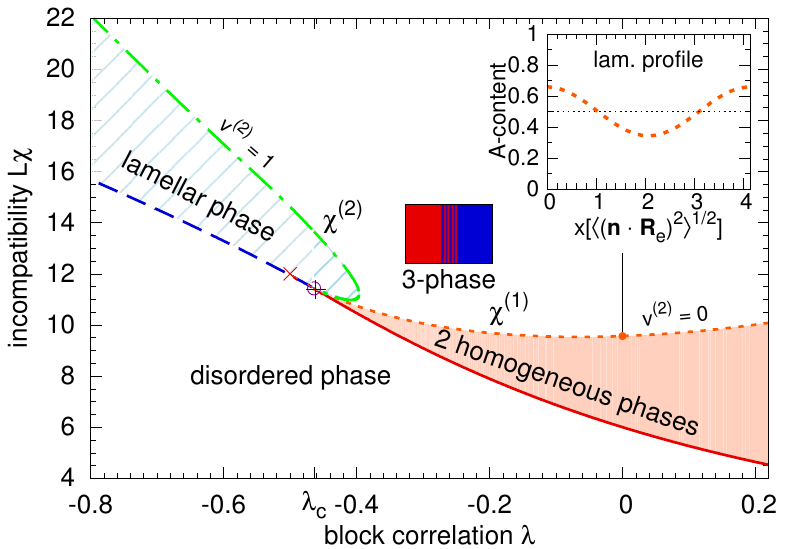}
\caption{(Color online)
Phase diagram for continuous triblocks with $k$-dependence of the free-energy functional eq.\ \xref{f_4th}. 
Line styles as in fig.~\ref{fig-chi2m-chih-M3}.
Crosses indicate the end points of the line of metastable macroscopic phase separation for $\lambda < \lambda_c$ ($\color{red}\times$) and of metastable lamellae for $\lambda > \lambda_c$ ($\color{blue}+$).
Top right inset: profile of $A$ fraction 
in lamellar shadow at $\lambda = 0$.
\label{fig-chi2m-chih-cont-k4}}
\end{figure}
The effects of the wave number variation within our fractionation scheme can be observed in fig.~\ref{fig-chi2m-chih-cont-k4}, for random continuous triblocks.
The boundary between global macroscopic phase separation and the three-phase region at $\lambda > \lambda_{c}$ is located at lower incompatibilities than that obtained with the simplified free energy (cf.\ fig.~\ref{fig-chi2m-chih-cont}). 
Global lamellar phase separation is found to be stable in a larger parameter region and to extend into the half-plane $\lambda > \lambda_{c}$. However, upon further increasing $\chi$ in the system with global lamellar phase separation at $\lambda > \lambda_{c}$, we find a reentrance into the fractionated three-phase coexistence.
Note that the amplitude of the lamellar shadow at the onset of fractionation attains a reasonably small value also at a block correlation distant from the critical one
(cf.\ the sinusoidal profile in fig.~\ref{fig-chi2m-chih-cont-k4}).

The main 
advantage of the lamellar free energy eq.~\xref{fmps_compl}
is the principal possibility of global lamellae also at $\lambda > \lambda_c$, since
the optimal wave number changes with increasing $\chi$ even in a fixed sequence distribution
(similar to the mechanism of global microphase separation invoked 
in ref.~\cite{leibler80}, which, however, considered one-component diblock copolymers only).
Global lamellar phase separation is found to follow the three-phase coexistence at high incompatibilities also within SCFT (see fig.~\ref{ph-diag-scft} below).

\begin{figure}[h]
\includegraphics[width=8cm]{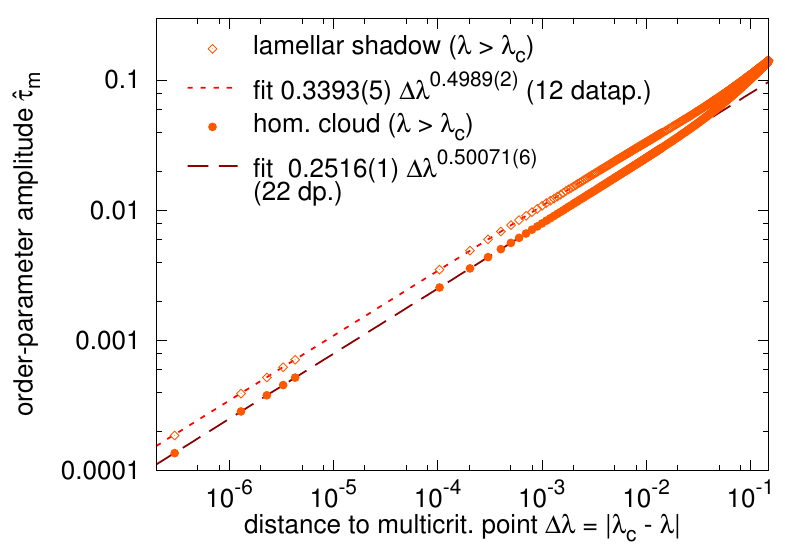}
\caption{\label{fig-ampl-lambda_contk4}(Color online) Order-parameter amplitude of homogeneous clouds and  lamellar shadow at the onset of fractionation for continuous triblocks, with $k$-dependent fourth order. 
}
\end{figure}

The scaling of the order-parameter amplitude on approach to the multicritical point 
is exctracted from the regularly shaped lamellar shadow line in fig.~\ref{fig-ampl-lambda_contk4}. 
Both the lamellar shadow and the macroscopic cloud amplitudes vanish with an exponent of $0.5$, corroborating the findings with the simplified lamellar free energy.


\section{Alternative approach: SCFT}
\label{chap_SCFT}
\subsection{Method}
An alternative method to determine the phase behavior of random triblocks employs self-consistent field theory (SCFT) \cite{helfand-scft, fred-book}. In order to analyze the phase coexistence of homogeneous and lamellar phases with finite volume fractions, it starts out from the grand-canonical partition function,
\ba
{\cal Z}_G &=& \sum_{ \left\{N_{\nu} \right\} } \prod_{\nu} 
\frac{1}{N_{\nu} !} \left( 
\frac{\varrho_0 \zeta_{\nu}}{L {\cal Q}_{\rm o}}\right)^{N_{\nu}}  \label{ZG}
\\ && \int \D\bm\left[r (s)\right]
\exp\left\{ 
- \cal{H}_{\chi} - \cal{H}_{\kappa}- \cal{H}_{\text{W}} \right\},
\nonumber\ea
with $\ln \zeta_{\nu}$ being the excess chemical potential of species $\nu$, $\nu=1, \ldots, 6$. 
${\cal Q}_{\rm o}$ denotes the configurational partition function (without translation) of a single, non-interacting Gaussian chain.  
Via the incompressibility demand (see below),
the sum over the sets $\{N_\nu\}$ of species numbers is restricted by the constraint, $\sum_{\nu=1}^6 N_\nu = N$. 
Therefore not all the fugacities $\zeta_\nu$ are independent, and we set $\zeta_{ABA} \equiv 1$. By virtue of the symmetry A$\leftrightharpoons$B of the $\lambda$-distribution and the coexisting phases, $\zeta_{AAA}=\zeta_{BBB}$, $\zeta_{AAB}=\zeta_{BBA}$, and $\zeta_{ABA}=\zeta_{BAB}$.

\begin{figure}
\includegraphics[width=\columnwidth]{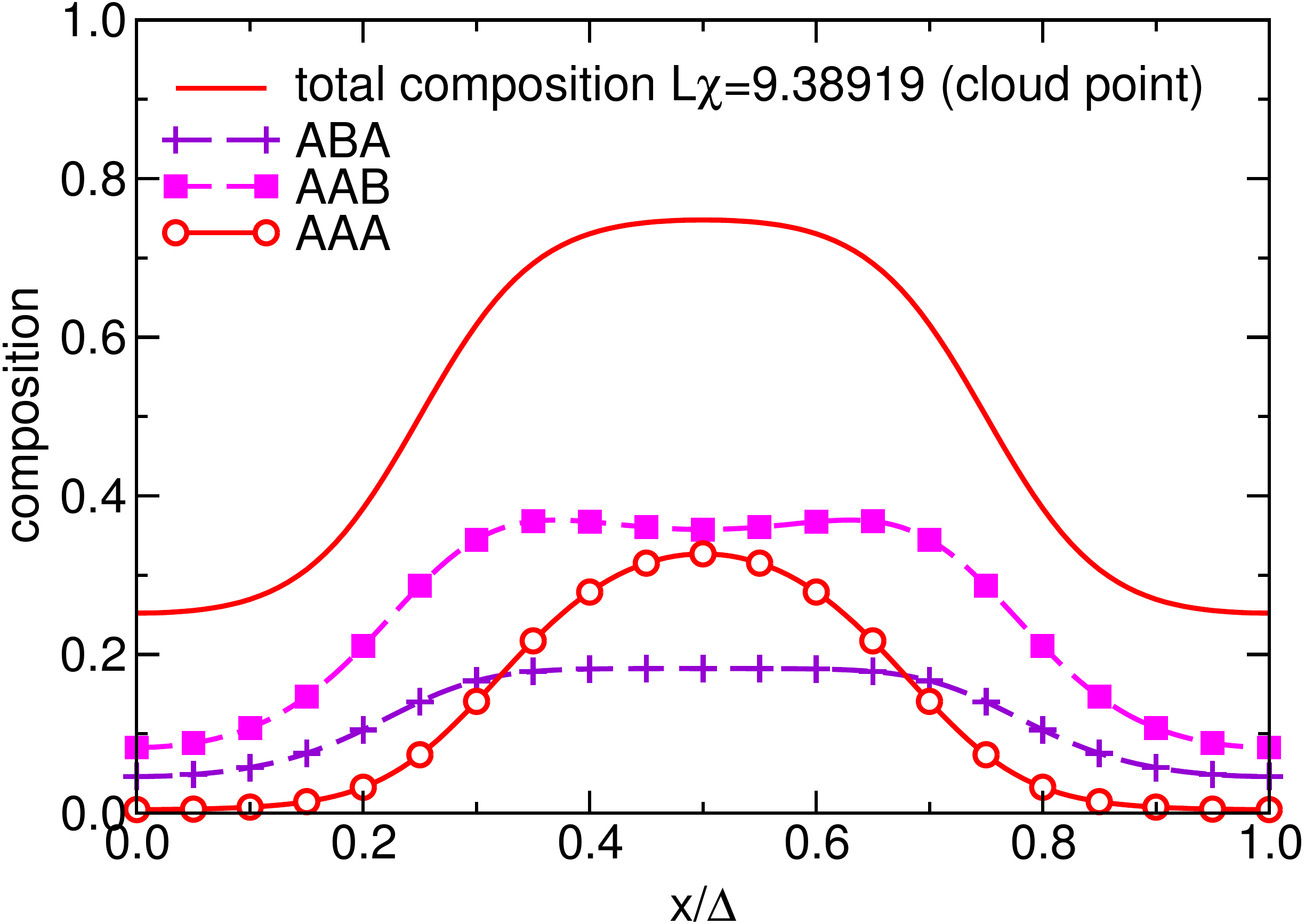}
\caption{(Color online) Local composition ($A$-segment density) 
profiles of the lamellar cloud phase at $\lambda=0$. 
Solid line: total density, 
symbols (key in the plot): due to one copolymer species. 
The spatial coordinate $x$ is normalized by the lamellar spacing, $\Delta=4.148 
R_{\rm e}$, where $R_{\rm e}$ denotes the rms end-to-end distance of a non-interacting triblock copolymer.
\label{profile0}}
\end{figure}

Similarly to the formalism in the previous sections, $A$- and $B$-density fields with their respective auxiliary fields $w_A$ and $w_B$ are introduced to decouple the interacting chains. 
The incompressibility constraint 
is accounted for by an additional Lagrange field $\xi$ and automatically imposes the constraint on the species numbers. Within the saddle point approximation, we obtain the excess grand-canonical potential, $g\equiv G/N+1$, per molecule:
\begin{equation}
g = \frac{1}{V} \int {\rm d}^3r\; \left\{ \chi L \phi_{A} \phi_{B} - w_{A} \phi_{A} - w_{B} \phi_{B}\right\},
\end{equation}
where the saddle point values of the fields and densities are determined by the self-consistent set of equations
\begin{subequations}
\begin{align}
\phi_{A}+\phi_{B} &= 1,  \\
w_{A} &= \chi L \phi_{B} + \xi, \\
w_{B} &= \chi L \phi_{A} + \xi, 
\end{align}
\begin{align}
\phi_{A}({\bm r}) &= -  \sum_{\nu=1}^6 \zeta_\nu V \frac{\delta  {\cal Q}_\nu}{\delta w_{A}({\bm r})}, \\
\phi_{B}({\bm r}) &= -  \sum_{\nu=1}^6 \zeta_\nu V \frac{\delta  {\cal Q}_\nu}{\delta w_{B}({\bm r})}.
\end{align}
\end{subequations}
The global concentration $p_{\nu}$ of species $\nu$ is given by 
\be 
p_{\nu} = \frac{1}{V} \int {\rm d}^3{r}\; \phi_\nu({\bm r}) = \zeta_{\nu} {\cal Q}_{\nu}. 
\ee 
The saddle point equations involve the partition functions, ${\cal Q}_{\nu}$, of single copolymer chains of species $\nu$ in the external fields $w_{A}$ and $w_{B}$:
\ba
&&\!{\cal Q}_{\nu}  = \left\langle\exp\left\{ - \int_0^{L} \d s\, w_{\nu} \left(\bm r(s), s\right)\right\}\right\rangle,\mbox{ with} \label{Q_nu} \\ 
&&\!w_{\nu} \left(\bm r(s), s\right) \mathrel{\mathop:} = \frac{1 + q_{\nu}(s)}{2} w_{A}(\bm r(s))
+ \frac{1 - q_{\nu}(s)}{2} w_{B}(\bm r(s)) 
\nonumber\ea
with the conformational average defined in eq.~\xref{wiener-av}. In the following, only the continuum limit of Gaussian chains is considered. For a structured phase, the ${\cal Q}_{\nu}$ and density profiles are expressed in terms of statistical weight propagators $q_{\nu}(\bm r, s)$, $q_{\nu}^{\dagger}(\bm r, s)$ along a Gaussian chain,
\begin{subequations}
\ba  
q_{\nu}(\bm r, s) &=& \left\langle \exp\left\{ - \int_0^{s} \d s' \, w_{\nu} \left(\bm r(s'), s' \right)\right\} \right\rangle_{ H_{\rm W}^{(s)\, (\bm r(s) = \bm r)} }
\\
q_{\nu}^{\dagger}(\bm r, s) &=& \left\langle \exp\left\{ - \int_s^{L} \d s' \, w_{\nu} \left(\bm r(s'), s' \right)\right\} \right\rangle_{ H_{{\rm W} \, (\bm r(s) = \bm r)}^{(L-s)} } 
\label{prop}\ea
\end{subequations}
where $H_{\rm W}^{(s)\, (\bm r(s) = \bm r)}$ and $H_{{\rm W} \, (\bm r(s) = \bm r)}^{(L-s)}$ are the conformation statistical weights for a chain of length $s$ having its end point at $\bm r$ and for  a chain of length $L - s$ having its start point at $\bm r$, respectively. The single-chain partition functions ${\cal Q}_\nu$ are calculated according to:
\be 
{\cal Q}_{\nu} = \frac{1}{V}\int\d^3 r\, q_{\nu}(\bm r, s) q_{\nu}^{\dagger}(\bm r, s),\qquad \forall s\in[0, L].
\ee
The propagators obey the modified diffusion equations:
\begin{subequations}
\ba
\left( \frac{\partial }{\partial s}  - \triangle_{\bm r}  + w_{\nu} \right) q_{\nu}(\bm r, s) &=& 0,\\
\left( \frac{\partial }{\partial s}  + \triangle_{\bm r}  - w_{\nu} \right) q_{\nu}^{\dagger} (\bm r, s) &=& 0.
\ea
\end{subequations}
These partial differential equations are solved via a spectral method \cite{matsenPRL94}. As a result, we obtain the equilibrium spacing and the free energy of the lamellar phase, 
as well as detailed composition (concentration) 
profiles of the different species in a lamellar domain. An example of a 
composition profile is shown in Fig.~\ref{profile0} for $\lambda=0$ at the lamellar cloud point, $L\chi=9.389\,19$.

The canonical free energy can be obtained via a Legendre transformation:
\begin{equation}
F = 
G + \sum_{\nu=1}^6 N_\nu \ln \zeta_\nu + N \ln \frac{\varrho_{\rm o}}{L {\cal Q}_{\rm o}}
\end{equation}
Thus the excess Helmholtz free energy $f$ per molecule takes the form
\begin{align}
f \equiv & \frac{F}{N} - \ln \frac{\varrho_{\rm o}}{L {\cal Q}_{\rm o}}   =  g - 1 + \sum_{\nu=1}^6 p_\nu \ln \zeta_\nu \nonumber\\
=& {\sum_\nu p_\nu (\ln p_\nu -1)} + {\frac{\chi L}{V} \int {\rm d}^3r\; \phi_{A} \phi_{B}} \\
& - {\sum_\nu p_\nu \ln {\cal Q}_\nu - \frac{1}{V}  \int {\rm d}^3r\; (w_{A} \phi_{A} + w_{B} \phi_{B}) }
\nonumber
 \end{align}
The first term corresponds to the entropy of mixing of the different species, 
the second term quantifies 
the free energy due to the repulsion of unlike monomer types, and the last two terms 
describe the loss of conformational entropy of the polymers in a spatially inhomogeneous environment.

\subsection{Three-phase coexistence lines and fractionation}
\begin{figure}
\includegraphics[width=\columnwidth]{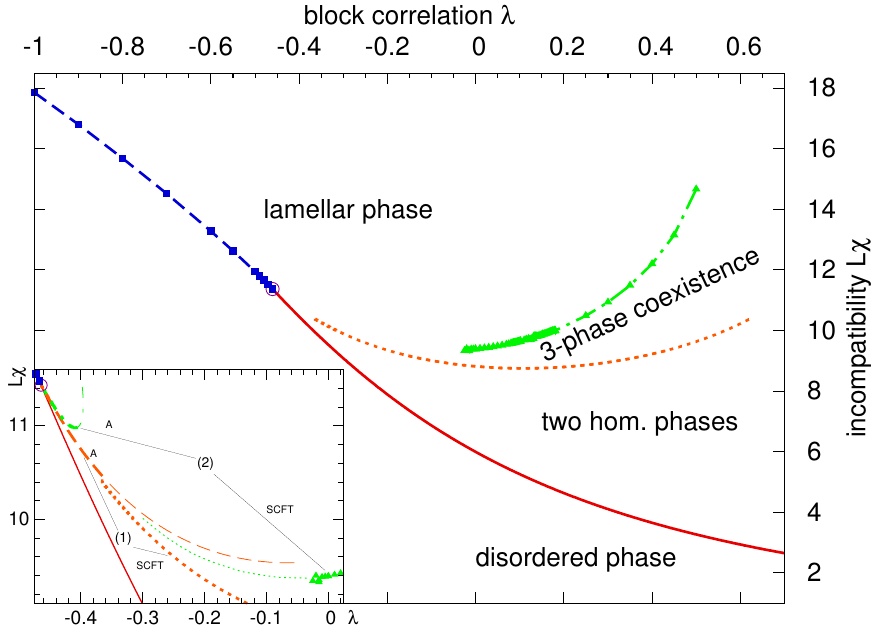}
\caption{(Color online) Phase diagram for continuous triblocks within SCFT. 
Line styles as in fig.~\ref{fig-chi2m-chih-M3}. 
Symbols highlight 
data:
triangles (green) lamellar cloud points, 
squares (blue) order-disorder transitions. 
The inset shows a detail, 
with the three-phase boundaries from the analytical method (A) added 
(cf.~fig.~\ref{fig-chi2m-chih-cont-k4}), in dashed (orange)
homogeneous, in dot-dashed (green)
lamellar cloud points. 
Thin dots (green) extrapolate the SCFT lamellar cloud points to the multicritical point.
\label{ph-diag-scft}}
\end{figure}

\subsubsection{Homogeneous cloud phases}
If we approach three-phase coexistence by increasing the incompatibility $L \chi$ from a low value at fixed $\lambda$, the lamellar phase (shadow) will emerge from the homogeneous, $A$-rich and $B$-rich phases (clouds) with an infinitesimal volume fraction. At the onset of three-phase coexistence, the sequence distribution of the two homogeneous cloud phases is a $\lambda$-defined one. 
In the grand-canonical ensemble, we determine the two independent excess chemical potentials, $\zeta_{AAA}$ and $\zeta_{AAB}$, of the cloud phases as to reproduce the composition of the $\lambda$-distribution. Since the incipient lamellar phase can exchange polymers with the cloud phases, its properties are calculated in the grand-canonical ensemble. To this end, we minimize the grand-canonical potential, $g$, at given $\zeta_{AAA}$ and $\zeta_{AAB}$ with respect to the lamellar period or spacing $\Delta$. The onset of three-phase coexistence occurs at the incompatibility, at which the so-minimized grand-canonical potential of the lamellae equals the grand-canonical potential of the cloud phases. 
The $(\lambda,\chi)$-points, 
at which the homogeneous, $A$-rich and $B$-rich phases are the cloud phases, 
are shown as a dotted curve in the phase diagram fig.~\ref{ph-diag-scft}. In the range $-0.17 < \lambda < 0.43$, the data were calculated with a spatial resolution of $32$ Fourier components, in the remaining range with $12$ components.

\subsubsection{Lamellar cloud phase\label{scft-lam-cloud}}
As we progress into the three-phase coexistence toward larger incompatibilities $L\chi$,
the volume fraction of the lamellar phase grows, 
while that of the homogeneous, $A$-rich and $B$-rich phases decreases. 
At the end of three-phase coexistence, the lamellar phase occupies the entire volume, and the homogeneous phases continuously disappear with a vanishing volume fraction. 
In order to determine this cloud point of the lamellar phase, 
we calculate the properties of the latter in the canonical ensemble, 
where its sequence distribution is fixed to $\lambda$-distribution. 
The canonical free energy $f$ is minimized with respect to the lamellar spacing $\Delta$. 
Then, the two independent excess chemical potentials for this optimal lamellar structure are measured, and the properties of the incipient homogeneous phases are calculated in the grand-canonical ensemble at the so-determined chemical potentials. 
Finally, $L\chi$ is adjusted such that the lamellar cloud and the incipient homogeneous shadow phases have the same grand-canonical potential at identical excess chemical potentials.
The resulting boundary points of three-phase coexistence toward large $L\chi$ are marked in fig.~\ref{ph-diag-scft} by triangles on a dot-dashed line. 
Twelve Fourier components were considered in this calculation.

\subsection{Phase coexistence with finite volume fractions}

The properties of 
a general 
fractionated state of three coexisting phases are computed in the grand-canonical ensemble. 
As in sec.~\ref{scft-lam-cloud}, we consider a lamellar phase (marked by the superscript $(2)$) with volume fraction $v^{(2)}$, coexisting with two homogeneous, $A$-rich and $B$-rich phases, with joint volume fraction $1-v^{(2)}$. 
A fractionated state with given volume fractions is located by simultaneously adjusting the two independent excess chemical potentials $\zeta_{AAA}$ and $\zeta_{AAB}$
and the incompatibility $L\chi$ 
such that the weighted sum of the sequence concentration $p_1^{(2)}$, respectively $p_2^{(2)}$, 
in the lamellar phase 
and $p_1^{(1)}$, respectively $p_2^{(1)}$, in the two homogeneous phases 
gives the global concentration of a $\lambda$-distribution (cf.\ eqs.~\xref{p_lambda} and \xref{constraint-p-lambda}),
and such that the grand-canonical potentials of all three phases are equal \cite{Buzzacchi06},
$g^{(1)} = g^{(2)}$.
In the limit $v^{(2)} \to 0$, we recover the cloud points of the homogeneous phases, 
in the limit $v^{(2)} \to 1$ we recover the cloud points of the lamellar phase. 
In contrast to the phases at their cloud points, 
none of the coexisting phases with finite 
volume fractions displays a $\lambda$-distribution 
(cf.\ sec.~\ref{sec-fract-distr} below). 

\begin{figure}[h!]
\includegraphics[width=8.1cm]{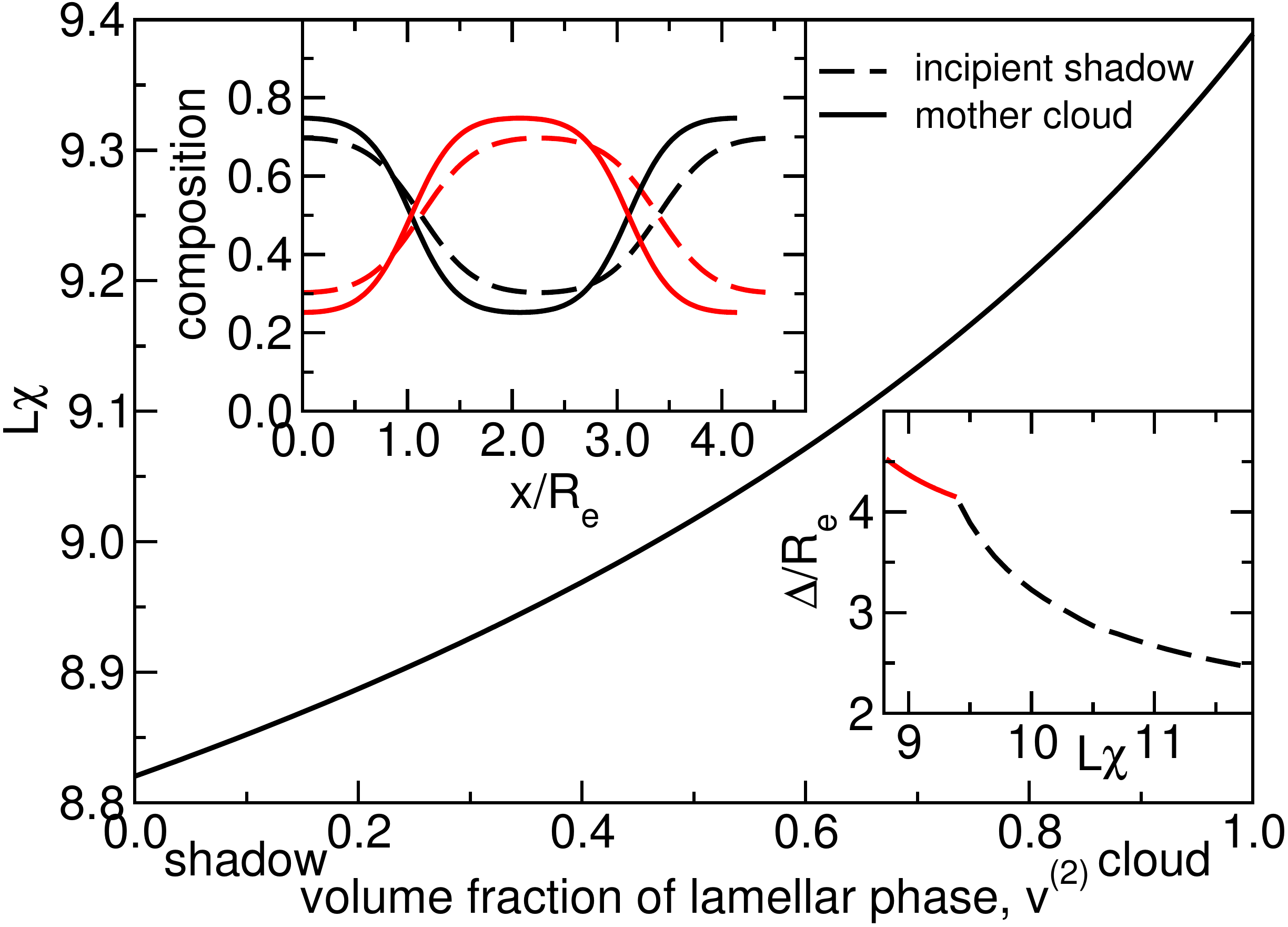}
\caption{(Color online) 
Incompatibility $L\chi$ against 
volume fraction $v^{(2)}$ of the fractionated lamellar phase at $\lambda=0$.
The end points of the curve mark the limits
of three-phase coexistence,
at which the lamellar phase is the incipient shadow, $v^{(2)}=0$, respectively the cloud, $v^{(2)}=1$. 
The top inset shows the lamellar $A$ density profiles in these limits. 
The bottom inset shows the spacing $\Delta$ of lamellae in the three-phase coexistence region and that of global lamellae for $L\chi>9.389\,19$.
\label{vol_frac_0}}
\end{figure}
The gradual change of the volume fraction of the lamellar phase upon increasing the incompatibility $L\chi$ at $\lambda=0$ is shown in fig.~\ref{vol_frac_0}. The inset presents the composition ($A$-segment density) profiles  of the lamellar phase at its shadow point $L\chi= 8.820\,43$ (dashed line), and at its cloud point $L\chi = 9.389\,19$ (solid line). 
We observe that 
the lamellar shadow's profile, though it is not confined to a single harmonic,
matches quite well the profile obtained
from the analytical method (see the inset in fig.~\ref{fig-chi2m-chih-cont-k4}), 
especially in the amplitude.
In contrast to results for one-component diblock copolymer melts \cite{leibler80}, but in agreement with predictions of random phase approximation, the lamellar spacing decreases upon increasing $L\chi$.

\section{\label{sec-fract-distr}Fractionated sequence distributions}

\begin{figure}[h!]
\includegraphics[width=6.2cm]{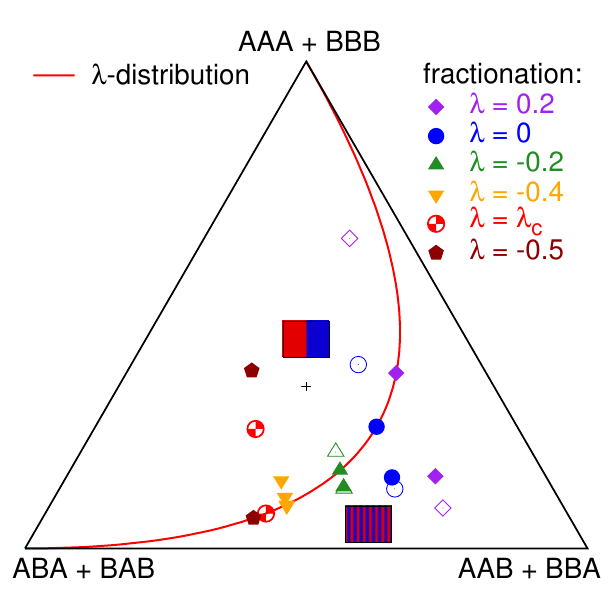}
\caption{(Color online)
Sequence distribution triangle for random continuous triblocks at various block correlations, 
with the extended analytical method, cf.\ the phase diagram in fig.~\ref{fig-chi2m-chih-cont-k4}. 
The diagram's center (${\scriptscriptstyle\bm+}$) corresponds to equal concentrations of all sequences ($p_{\nu}=\nicefrac{1}{3}$, $\nu=1,2,3$). 
One sequence distribution is represented as a linear combination of the vectors pointing from the center to the corners, each vector scaled with the 
concentration deviation $3\left( p_{\nu} - \nicefrac{1}{3} \right)/2$.
Distributions defined by $\lambda$ lie on the (red) curve, with $\lambda$ ranging from $-1$ at the triangle's bottom left corner to $+1$ at its top. 
Solid symbols on this curve mark the sequence distribution of the cloud phase(s) 
at the boundary line(s) of three-phase coexistence. Off-curve solid symbols mark the distributions of the coexisting shadow phases. Open symbols display the distributions of the coexisting states at equal volume fractions ($v^{(2)} = 0.5$).
\label{fig-conc_triangle}
}
\end{figure}

\begin{figure}
\includegraphics[width=6.2cm]{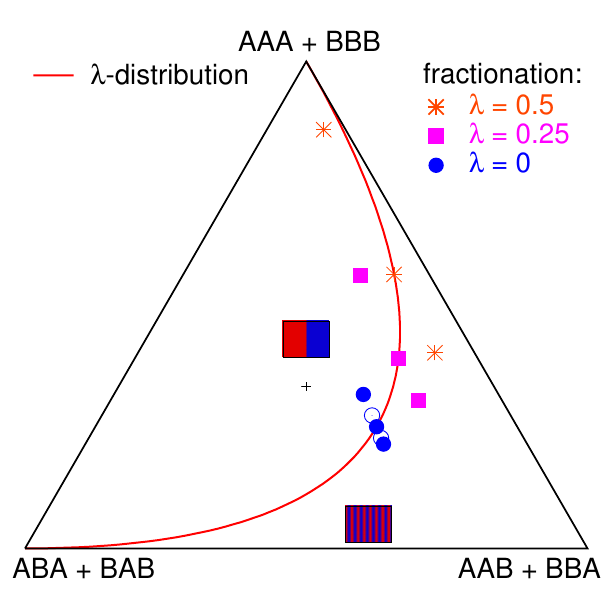}
\caption{(Color online) Sequence distribution triangle for random continuous triblocks within SCFT,
with the distributions of the coexisting phases at the beginning and at the end of three-phase coexistence for $\lambda=0$ (blue circles), $0.25$ (magenta squares), and $0.5$ (red stars), respectively.  For $\lambda=0$, open circles mark the distributions of the coexisting phases at $L \chi = 9.017\,23$, where the lamellar phase comprises half of the volume. The solid line represents $\lambda$-distributions.
\label{triangle}}
\end{figure}

In this section, we invoke both the analytical and the SCFT method to obtain detailed 
sequence distributions, which show the fractionation or sequence partitioning according to the coexisting phases' structures in random continuous triblocks.
In figs.~\ref{fig-conc_triangle} and \ref{triangle}, 
the sequence distributions of the coexisting phases are presented 
by means of composition triangles: 
Each corner represents one of the sequence classes defined in eq.~\xref{species-def}, 
a point within the triangle one sequence distribution.
Due to the A$\leftrightharpoons$B exchange symmetry of species combined into 
one sequence class, 
the distributions of
the two homogeneous phases within a macroscopically phase-separated state coincide in this triangle.

In fig.~\ref{fig-conc_triangle},  we present the fractionated distributions obtained by the analytical method with the restored $k$-dependence of the lamellar free energy. 
The sets for three supercritical values of the block correlation, 
$\lambda > \lambda_c$, 
$\lambda=0.2$ (diamonds), $\lambda=0$ (circles), and $\lambda = -0.2$ (up triangles),
visualize the following fractionation mechanism:
On the curve of $\lambda$-distributions, the solid symbol indicates the sequence
distribution of the homogeneous cloud phase(s) at the onset of fractionated three-phase coexistence. 
The solid symbol of the same shape and color to the bottom right of the curve, 
marks the distribution of the coexisting lamellar shadow phase (with zero volume fraction).
The finite deviation of the lamellar shadow's sequence distribution from the $\lambda$-distribution 
shows that the transition to three-phase coexistence is discontinuous.  
Upon increasing incompatibility, the lamellar phase's volume fraction increases (cf.\ fig.~\ref{fig-v2-Nchi-cont}), 
and its sequence distribution
departs ever more from the $\lambda$-distribution (the open symbols to the
bottom right of the $\lambda$-curve display lamellae at $0.5$ volume fraction). 
Sequence class 2 ($AAB$/$BBA$) substantially accumulates in the lamellar phase,   
also class 3 ($ABA$/$BAB$).  
Moreover, since the ratio of these
two sequence concentrations differs from the $\lambda$-defined ratio $p_{2}(\lambda)/p_{3}(\lambda)$,
the fractionated sequence distribution in the lamellar phase 
does not ensue from merely expelling homopolymers into the coexisting homogeneous phases
at a constant ratio of the other two sequence classes.
As the volume fraction of the homogeneous, initial cloud phase(s) decreases, 
their distribution (at volume fraction $0.5$ marked by open symbols to the top left)
deviates increasingly from the $\lambda$-curve, 
showing in turn a particular depletion in $AAB$/$BBA$ sequences. 

For $\lambda=-0.4$, the reentrant behavior of the
three-phase boundary line, cf.\ fig.~\ref{fig-chi2m-chih-cont-k4}, 
gives rise to various coexisting distributions (down triangles).
Upon increasing $\chi$ in the two homogeneous phases, the three-phase region appears with a lamellar shadow (nearly on the curve of $\lambda$-distributions, shift to the bottom hardly visible), which grows with $\chi$ in volume fraction until it becomes a lamellar cloud (now the symbol on the curve of $\lambda$-distributions) coexisting with two homogeneous shadows (triangle shifted slightly to the top). The homogeneous shadows are homopolymer-enriched and depleted in alternating sequences.
At an even higher $\chi$, the global lamellar phase gives way to a three-phase coexistence again.
The topmost triangle 
represents the distribution of the homogeneous shadows at this reentrance.  
For $\lambda \leq \lambda_c$, this lamellar cloud line is the only three-phase boundary.  
The topmost symbols for the critical and subcritical correlations
$\lambda = \lambda_c$ and $\lambda = -0.5$
show the distributions of the coexisting
homogeneous shadows which deviate markedly from the $\lambda$-distributions.  


In the distribution triangle of fig.~\ref{triangle}, we present the SCFT results for the sequence distributions of the coexisting phases for $\lambda=0$, $0.25$, and $0.5$. 
Again, the distributions of the cloud phases are represented by solid symbols 
on the solid curve of $\lambda$-distributions. For each value of $\lambda$, the distributions at the beginning and the end of three-phase coexistence are shown. At the lower incompatibility, the homogeneous phases are the clouds, and the coexisting lamellar shadow corresponds to the respective solid symbol shifted to the lower right corner, with its distribution enriched in $AAB$/$BBA$ sequences. At the higher incompatibility, the lamellar phase occupies the total volume, $v^{(2)}=1$,
and its distribution is represented by the cloud symbol on the $\lambda$-curve.
The distribution of the coexisting homogeneous shadows corresponds to the symbol shifted to the upper left side of the triangle. Two open circles mark the distributions
for equal volume fractions of the macroscopic and the lamellar phase-separated state, $v^{(2)}=0.5$,
at $\lambda=0$. In this situation, none of the coexisting phases is characterized by a $\lambda$-distribution; the homogeneous phases are rich in homopolymers, while alternating sequences segregate into the lamellar phase.
In comparison to the analytical results for the distributions at $\lambda=0$, apart from the qualitatively different feature of a lamellar cloud at higher incompatibilities, 
the sequence fractionation is found to be weaker. Note, however, the smaller transition incompatibilities to three-phase coexistence within SCFT,  which also result in smaller order-parameter amplitudes.

\begin{figure}
\includegraphics[width=\columnwidth]{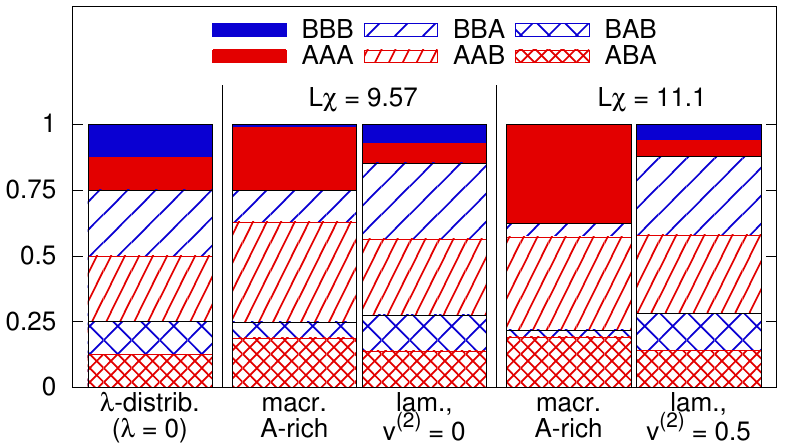}
\caption{\label{fig-fract-conc-k4}(Color online)
Detailed sequence distributions of the coexisting phases at $\lambda=0$ from analytical method (sec.~\ref{sec-frac-compl}), cf.~fig.~\ref{fig-conc_triangle}.
Leftmost chart: $\lambda$-distribution of the disordered melt.
Pairs of charts: distributions of the $A$-rich, homogeneous  
and the lamellar phase; 
left: at the onset of three-phase coexistence; 
right: at a lamellar volume fraction of $0.5$.
}
\end{figure}

\begin{figure}
\includegraphics[width=\columnwidth]{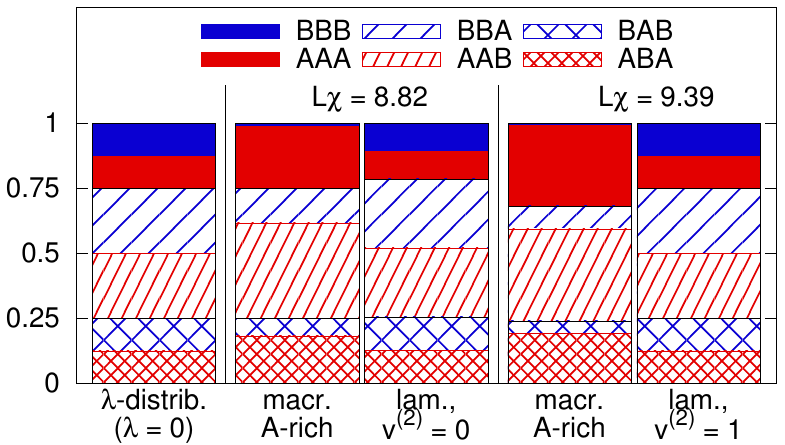}
\caption{(Color online) 
Detailed sequence distributions of the coexisting phases at $\lambda=0$ obtained with SCFT. 
Pairs of charts: distributions of the $A$-rich, homogeneous and the lamellar phase; 
left: at the onset; right: at the end of three-phase coexistence.
\label{boxchart}}
\end{figure}
Detailed sequence distribution diagrams for lamellar and macroscopic phases at $\lambda=0$ are
displayed in fig.~\ref{fig-fract-conc-k4} for the analytical method, and in fig.~\ref{boxchart} for SCFT.
The representation of all six species' concentrations additionally visualizes the segregation within a sequence class into $A$- and $B$-rich subspecies between the two homogeneous phases, 
which allows for an estimate of 
the macroscopic $A$ excess amplitude at different stages of fractionated three-phase coexistence. 
Due to the homogeneous phases' $A\leftrightharpoons B$ exchange symmetry, 
only the distribution of the $A$-rich, homogeneous phase is shown.
(The chart for the $B$-rich phase looks the same as the one depicted for the $A$-rich phase, only with letters $A$ and $B$ exchanged in the key.) 
The distributions obtained by both methods agree well.
While both diagrams reveal the preference of the fractionated lamellar phase for $AAB$/$BBA$ sequences, the accumulation is more distinctive 
in the analytical results, already at the onset of fractionation (cf.\ the central charts). 
Corresponding to the higher onset incompatibility, 
the macroscopic segregation into $A$- and $B$-rich subspecies 
also is at a more advanced stage.

\section{Discussion\label{chap-discussion}}

\subsection{Analytical mean-field approach\label{amf}}

The analytical mean-field theory 
is restricted in its validity, whenever a lamellar phase is addressed,
to small lamellar order-parameter amplitudes or
to the proximity of a continuous microphase transition.
In any case, 
it is able to analyze accurately and 
in detail the vicinity of the multicritical point $\left(\lambda_c, \chi_c\right)$, 
whose quality is found to depend sensitively on the number $M$ of segments per block. 
For a small number of segments per block ($M<7$), 
the wave number of the global ordered state grows continuously from zero, 
when decreasing $\lambda$ from $\lambda_c$.
A reentrance into 
the fractionated three-phase coexistence 
is observed for $\lambda \lesssim \lambda_c$. 
The critical exponent for the lamellar order-parameter amplitudes on approach to 
$\left(\lambda_c, \chi_c\right)$ is $1$, along both three-phase coexistence boundaries. 
For more segments per block ($M\geq 7$), the
structure factor of the $\lambda$-distribution develops a second peak
at finite $k$, such that the wave number of the global ordered state 
is discontinuous at  $\lambda_c$. 
Contrasting with the case $M<7$, 
we find a critical exponent of $0.5$ for the amplitudes of both lamellar and homogeneous phases,
along both three-phase boundaries. 
This behavior 
might be due to the intersection of transition lines to metastable, global ordered phases at 
multicriticality.

With 
the simplest version of the free-energy functional, eq.~\xref{f_4thk0}, 
a (global) lamellar cloud phase with $\lambda$-defined concentrations
can occur only at $\lambda<\lambda_c$ 
(for our system, a result qualitatively different from the SCFT predictions;
see sec.~\ref{disc-scft} below).
An enhanced version of our theory abandons this restriction 
by restoring the wave-number dependence of the quartic vertices 
in the lamellar free-energy function, eq.~\xref{fmps_compl};
see the location of the lamellar cloud boundary for random continuous triblock copolymers in
fig.~\ref{fig-chi2m-chih-cont-k4}. 
The critical exponent of $0.5$,
found for the order-parameter amplitudes 
along three-phase coexistence lines with the simplified theory, 
is corroborated by the enhanced analytical theory. 

With the complete wave-vector dependence of eq.~\xref{f_4th},
at fixed $\lambda$, a global lamellar phase can attain a lower free energy 
than global macroscopic phase separation at an incompatibility
$\chi_{\text{L}} > \chi_{\text{h}}$ -- a mechanism of microphase separation
proposed by Leibler and co-workers~\cite{leibler80,fred92}. 
Via our parameterization of a fractionated three-phase coexistence,
we take into account more degrees of freedom and 
find, instead of this mechanism, a 
refined competition to be effective: 
A structured phase first becomes stable in a subsystem with vanishing
volume fraction and with a sequence distribution different from the
global one. This onset of three-phase coexistence indeed occurs at a
smaller incompatibility $\chi < \chi_{\text L}$ than that of the global microphase
separation conjectured by Leibler.

\subsection{Numerical SCFT\label{disc-scft}}

The SCFT method invokes the mean-field approximation, too, 
but avoids the assumption of small order-parameter amplitudes and the single-harmonic approximation for the lamellar phase.
Thus, it
provides appropriate mean-field predictions for large regions of the phase diagram, 
but due to numerical problems fails 
as
the multicritical point is approached and both 
wave numbers and free-energy differences decrease.
Moreover, numerical SCFT 
is restricted to a small number of different components, 
and consequently allows us to address random copolymers with a
small number of blocks $Q$ only, 
which led to the choice $Q=3$ in this study. 
The SCFT calculation for random continuous block copolymers with $Q=3$ 
locates the entire three-phase region in the half-plane 
$\lambda>\lambda_c$ of the $\lambda$-$\chi$ diagram.

\subsection{Combining the results\label{comparison}}

Beyond the mean-field approximation, the analytical approach and SCFT
have different additional limitations, such that their results
for the location of three-phase boundaries are complementary: 
The analytical approach 
assumes the lamellar order-parameter amplitudes to be small, 
which is accurate in the vicinity of the multicritical point. 
In this region, however, 
also the free-energy differences between competing states 
(global lamellae, three-phase coexistence, two homogeneous phases) 
become minuscule (cf.\ the inset of fig.~\ref{fig-chi2m-chih-cont}),
which poses numerical difficulties for the SCFT calculations.
Hence, there is no regime where both approaches are simultaneously reliable, 
and a direct comparison is difficult. 
In the inset of fig.~\ref{ph-diag-scft}, we try to combine their results for the phase diagram of random continuous triblocks to one picture. 
The predictions for the cloud points of the homogeneous phases 
obtained by SCFT (dotted) and by the analytical method (dashed) match quite well, 
whereas the agreement for the cloud points of the lamellae is less satisfactory. 
Numerical SCFT results for these points (solid triangles) do not extend below $\lambda=-0.025$ 
due to the mentioned subtle free-energy differences in this region which control the phase behavior. 
The thin dotted line has not been computed, but marks our tentative 
extra\-polation of SCFT data toward the multicritical point,
based on the slope of
the lamellar cloud line determined with the analytical theory (dot-dashed) 
in the part that is in qualitative accordance (thick).  
The analytical prediction for this line (cf.~fig.~\ref{fig-chi2m-chih-cont-k4})
is 
enhanced 
relative to 
the rougher 
description presented in fig.~\ref{fig-chi2m-chih-cont}; 
cf.\ sec.~\ref{sec-frac-compl}. 
Still, owing to the delicate free-energy balance, 
the shape of this boundary line is bound to be more sensitive to
the approximation of small lamellar amplitudes in the theory  
than that of the other three-phase boundary, 
at which the lamellar phase is the shadow and all amplitudes are smaller. 
%

\section{Conclusions and Outlook\label{chap-conclusion}}

The analytical method and the numerical SCFT constitute complementary approaches, 
which both have their virtues and together provide a comprehensive mean-field
picture of the complex phase behavior of random triblock copolymers.
With both methods, we consistently reveal an extended three-phase coexistence region of
macroscopic and microscopic phase separation in random triblock copolymers, 
as suggested by simulations \cite{mmueller04}.  
Also, we discover 
the coexisting phases to select sequences that match their morpho\-logy.
Upon entering the three-phase region, the incipient shadow phase
emerges with vanishingly small volume fraction and with a sequence
distribution that already differs from the $\lambda$-distribution of the cloud phase.
Fractionation demixes the initial random (here Markovian)
distribution into sequence classes (following the analytical approach, progressively),
a separation mechanism which might prove useful to isolate wanted species
in polymer blends.

Our analysis has been restricted to mean-field theory. 
For the macroscopic phase separation of the disordered state 
at $\lambda >\lambda_c$, the critical 
region $(\chi - \chi_{\text h})$, within which the mean-field approximation fails,
has been estimated with the help of a Ginzburg criterion~\cite{houd-mmuell02}. 
The latter yields a Ginzburg number $\text{Gi} \propto Q^2/M$, i.e., the
critical region does not shrink simply with chain length $QM$, in contrast to
naive expectation. For fixed $Q$, such as considered here, the mean-field
predictions are correct in the limit of large $M$. The transition from
the disordered to a global microphase-separated state ($\lambda<\lambda_c$),
is expected to be weakly first-order due to
fluctuations~\cite{brazov75,fred-helf87}.  For the transition lines to
three-phase coexistence and the multicritical point at $\lambda
=\lambda_c$, the effects of fluctuations remain to be
explored. Whereas for simpler phase diagrams, it has been shown that
the Lifshitz point at $\lambda =\lambda_c$ is destroyed by
fluctuations, the situation here is more complicated due to the fact
that \emph{four\/} phase states meet in the multicritical point.

Phase coexistence enabled by component selection 
might be of interest for various other multi-component systems, 
cf., e.g., refs.~\cite{sollich02,fasolo-sollich04}.
Specifically for polydisperse copolymers, sequence fractionation
can be generalized starting from the case considered here:
A straightforward extension is to study random block copolymers 
asymmetric in global $A$-/B-content, 
which apart from the lamellar state, display other structured ordered morphologies, 
such as spheres on a bcc lattice or hexagonally arranged cylinders \cite{leibler80,potemk-panyuk98}.
Other generalizations would include copolymers either with an arbitrary number of blocks or 
built from more than two segment types. 
Fractionation may also give rise to structured phases beyond the ordered microphases.
Particularly promising in this context are random copolymers with
many blocks, which might display
frozen, random structures 
in coexistence with macroscopically phase-separated states.

\begin{acknowledgments}
We thank Christian Wald for 
valuable advice. Funding by the Deutsche Forschungsgemeinschaft through Grants No.
SFB-602/B6 and No. Mu1674/9 is gratefully acknowledged.
\end{acknowledgments}

\section*{Appendix}
\appendix
\section{\label{app_wiener-av}Gaussian-chain averages}
Equation \xref{rho_sp} and vertices of the expansion eq.~\xref{f_4th} contain
$n$-point 
correlations of the Gaussian-chain measure:
\begin{align} 
\lefteqn{ 
\left\langle \exp\Biggl\{
- i \sum_{\text r =1}^{n 
}\bm k_{\text r} \cdot
    \bm r(s_{\text r}) 
    \Biggr\}\right\rangle = \label{exp-w-av}
}\\ &
\delta_{\sum_{\text r}\bm k_{\text r}, \bm 0}
\exp\left\{
\sum_{\text r < \text r\prime}|s_{\text r}-s_{\text r\prime}|\bm k_{\text r}\cdot\bm k_{\text r\prime} 
\right\} 
\nonumber\end{align}
(
derived for continuous chains in \cite{gcz96}, appendix B).

\section{\label{app_vertexfcts}Vertex functions and 
moments}
In eqs.~\xref{rho_sp} and \xref{f_4th}, we also introduced the following functions:
The discrete Debye function 
\begin{align} \label{debye-L}
D (L, k^2) \mathrel{\mathop:}= 
&\sum_{s_1,s_2=1}^L 
\left\langle 
\e^{ -i \left( \bm k_1\cdot \bm r(s_1) + \bm k_2 \cdot \bm r(s_2) \right) } 
    \right\rangle
\\ =& \sum_{s_1,s_2=1}^L
\e^{ \left| s_{2} - s_{1}\right| \bm k_{1}\cdot\bm k_{2}}
\,\delta_{- \bm k_{2}, \bm k_1 =\mathrel{\mathop:} \bm k}
\nonumber\\
=&
\frac{L(1+\e^{-k^2})}{1-\e^{-k^2}}
-\frac{2\e^{-k^2}(1-\e^{-Lk^2})}{(1-\e^{-k^2})^2},
\nonumber
\end{align}
and the 
structure factors, for individual sequences:
\begin{align}
S_{\nu}(k^2) \mathrel{\mathop:}= 
&\sum_{s_1,s_2=1}^L q_{\nu}(s_1)q_{\nu}(s_2)
\left\langle\e^{-i \left( \bm k \cdot \bm r(s_1) + \bm k_2 \cdot \bm r(s_2)
    \right)}\right\rangle \nonumber\\
= &\sum_{s_1,s_2=1}^L
q_{\nu}(s_1)q_{\nu}(s_2)
\e^{-k^2|s_2-s_1|},
\label{S_k}
\end{align}
\be\label{S_2}
S^{(\alpha)}_{\nu}(\bm k_1,\bm k_2) \mathrel{\mathop:}= 
\sum_{s_1,s_2,s_3=1}^L
q_{\nu}(s_1)q_{\nu}(s_2)
\left\langle\e^{-i\sum_{\text r=1}^3 \bm k_{\text r} \cdot \bm r(s_{\text r})
}\right\rangle,
\ee
\begin{align} \label{S_4}
\lefteqn{
S^{(\beta)}_{\nu}(\bm k_1,\bm k_2,\bm k_3)
\mathrel{\mathop:}= 
}\\& 
\sum_{s_1,s_2,s_3,s_4=1}^L
q_{\nu}(s_1)q_{\nu}(s_2)q_{\nu}(s_3)q_{\nu}(s_4)
\left\langle
\e^{-i \sum_{\text r=1}^4 \bm k_{\text r} \cdot \bm r(s_\text r) 
}
\right\rangle,
\nonumber\end{align}
\begin{align}\label{S_5}
\lefteqn{
S^{(\gamma)}_{\nu}(k_1^2, k_2^2)\mathrel{\mathop:} = 
}\\&
\sum_{s_1,s_2,s_3,s_4=1}^L
q_{\nu}(s_1)q_{\nu}(s_2)q_{\nu}(s_3)q_{\nu}(s_4)
\e^{- k_1^2 |s_2- s_1| - k_2^2 |s_4 - s_3|}.
\nonumber\end{align}
Again, the length scale is the 
effective segment length $b$,
$k^2 \mathrel{\mathop:}= b^2 \tilde k^2/(2d)$, 
with $\tilde k$ the physical wave number.

For the global $\lambda$-distribution, 
the type correlation of two monomers on the same chain, 
whose block numbers differ by 
$\Delta\beta(s_1, s_2) \in\left\{0,\ldots, Q-1\right\}$, can
be calculated directly via the transition matrix $\hat M$ \xref{t-matrix}:
\begin{align}
\label{q-corr-lambda} 
\left[q(s_1) q(s_2)\right]_{\lambda} &
\mathrel{\mathop:}= \sum_{\nu} p_{\nu}(\lambda) q_{\nu}(s_1) q_{\nu}(s_2) \nonumber\\&
= \lambda^{|\Delta\beta(s_1, s_2)|}. 
\end{align}
Summing over all monomer pairs gives the second-order 
moment (cf.~eq.~\xref{nth-mom}) for a $\lambda$-distribution:  
\be\label{q_2_nat} \begin{split}
m_{2} (\lambda)
&\mathrel{\mathop:}=
\frac{1}{L^2}\sum_{s_1,s_2=1}^L\left[q(s_1)q(s_2)\right]_{\lambda}
\\  
   &\overset{Q=3}{=} \frac{1}{3} + \frac{2\lambda}{9}\frac{
(2- 3\lambda + \lambda^3)}{(1-\lambda)^2}.
\end{split}
\ee
Inserting eq.~\xref{q-corr-lambda} into eq.~\xref{S-p-nu} and performing the sum over all pairs
yields the expression eq.~\xref{S-k-nat} for the second-order structure factor $S\left(k^2\right)$ in a $\lambda$-distribution.
We abstain from presenting within this paper our computations 
of the structure factors, eqs.~\xref{S_k}--\xref{S_5},
of a $\lambda$-distribution for general $Q$
(the expression eq.~\xref{S-k-nat}
had been given earlier in~\cite{wald_diss}), 
and of individual sequences for $Q=3$.
Obtaining the lengthy expressions for the fourth-order structure factors 
requires extended sorting of the multiple sums' terms due to combinatorics.

\section{\label{app_crushed-polymers}Macroscopic phase separation}
Within the `crushed polymer picture', we derive for the free energy 
of coexisting homogeneous phases a closed expression that is not limited to small order-parameter amplitudes.
Here, each chain reduces to
one structureless particle 
with an $A$ excess $\tilde q_j$ equal to
the average over all segments on that chain.
Again with a field-based approach,
the calculation of the free-energy functional 
is analogous to that in sec.~\ref{sec:fmps}, 
but simpler, since conformational averages are obsolete for only one position $\bm r_j$ per chain.
(For a replica-based derivation 
see \cite{wald_diss}; 
the results prove to agree with Flory-Huggins theory \cite{flory53}.) 

For $Q$-block copolymers, it is sufficient to distinguish $(Q+1)$ components
according to their 
A excess:
\be 
\tilde q_l \mathrel{\mathop:}= \frac{2l-Q}{Q} = - \tilde q_{Q-l},\quad
l\in\left\{0,1,\ldots,Q\right\}.
\label{crush_poly}
\ee
In the case of symmetric triblock copolymers, the four component probabilities $\tilde p_{l}$ 
are related to the sequence probabilities defined in eq.~\xref{p_lambda} via
\be
\tilde p_{0} = \tilde p_{3} = \frac{p_1}{2}, \quad\tilde p_{1} = \tilde p_{2} = \frac{p_2 + p_3}{2} = \frac{1 - p_1}{2}.
\ee
With the coarse-grained component densities
\be \varrho_{l} (\bm r) = L\sum_{j=1}^N\delta_{\tilde q_j, \tilde q_l} \delta(\bm r - \bm r_j),\quad
l\in\left\{0,1,\ldots,Q\right\},
\label{def-rho_l}\ee
the total and $A$ excess densities are
\be\label{densities}
\varrho(\bm r) = \sum_{l} \varrho_l(\bm r) \text{ and }
\sigma(\bm r) = \sum_{l} \tilde q_{l} \varrho_{l}(\bm r),
\ee
and the partition function to calculate is 
\begin{align}
{\cal{Z}} = &\prod_{j=1}^{N} \left( \int \frac{\d^d r_j}{V} \right)
\\
&
\exp\left\{ 
\frac{1}{4 \varrho_0}\int\d^d r\, 
\biggl( 
\chi\Bigl( \sigma (\bm r) \Bigr)^2 - 2\kappa\Bigl(\varrho(\bm r) 
\Bigr)^2 
\biggr)
\right\}
\nonumber\end{align}
Introduction of additional fields, similarly as in eq.~\xref{part-fct}, 
and elimination of the original fields at the saddle point, 
yields the effective Hamiltonian per chain
\begin{align}
\lefteqn{ \tilde h = }\\
& \frac{1}{4N\varrho_0} \int\d^d x  \biggl( 
\chi\Bigl( \hat\tau (\bm x) \Bigr)^2 - \frac{2}{\kappa}
\Bigl(\hat\omega(\bm x) 
\Bigr)^2 
\biggr)
- \sum_{l} \tilde p_{l} \ln \tilde z_{l},
\nonumber\end{align}
with the single-component partition functions
\be
\tilde z_{l} \mathrel{\mathop:} =  \frac{1}{V} \int\d^d x \exp
\left\{ \frac{L}{2\varrho_0} 
\biggl( 
\chi \tilde q_{l} \hat\tau(\bm x) - 2\hat\omega(\bm x)
\biggr)
\right\}.
\ee
The general ansatz of $K\leq (Q+1)$ homogeneous phases,
\be \nonumber 
\left. 
\begin{array}{l} 
\hat\omega(\bm x) 	= \hat\omega^{(k)}  = \sum\limits_{l} \hat\omega_{l}^{(k)} 
,\\
\hat\tau(\bm x) 	= \hat\tau^{(k)} = \frac{1}{\kappa} \sum\limits_{l} \tilde q_l \hat\omega_{l}^{(k)},
\end{array}
\right\}
\bm x\in V_{\text{h}}^{(k)},\, k\in\left\{ 1,\ldots, K\right\},
\ee
with volume fractions $v^{(k)} \mathrel{\mathop:} = |V_{\text h}^{(k)}|/V$ gives
\begin{align}
\tilde h & = 
\frac{L}{4\varrho_0^2} \sum_k v^{(k)} 
\Bigl( \chi(\hat\tau^{(k)})^2
- \frac{2}{\kappa} (\hat\omega^{(k)})^2
\Bigr)
- \sum_{l}\tilde p_{l} \ln \tilde z_{l},\nonumber\\
\tilde z_{l} & = \sum_k v^{(k)} \exp\left\{ 
\frac{L\left( \chi\tilde q_l \hat\tau^{(k)} - 2\hat\omega^{(k)}\right)}{2\varrho_0}  
\right\}.
\label{funct-hom}
\end{align}
We optimize $\tilde h$ with respect to the $\{ v^{(k)}, \hat\omega^{(k)}, \hat\tau^{(k)} \}$,
with Lagrange multipliers $\Lambda_1$, $\Lambda_2$, $\Lambda_3$
for the constraints of number conservation, $\sum_{k} v^{(k)} = 1$,
constant global density, $\kappa\varrho_0 = \sum_{k} v^{(k)} \hat\omega^{(k)}$, 
and $A$ excess, here $\sum_{k} v^{(k)} \hat\tau^{(k)} = 0$. 
Solving the equilibrium conditions for nearly incompressible density conjugates,
\be \frac{ \hat\omega^{(k)} }{\kappa\varrho_0} = 1 + C^k \kappa^{-1} + {\cal{O}}\left(\kappa^{-2}\right),\ee
we find expressions for the density-conjugate differences
\be 
\lim_{\kappa\to\infty} \frac{\hat\omega^{(k)} - \hat\omega^{(k')}}{\varrho_0} 
= \frac{\chi}{4}\frac{(\hat\tau^{(k)})^2 - (\hat\tau^{(k')})^2 }{\varrho_0^2}, 
\ee
quadratic in the $A$-excess conjugates, such as in eq.~\xref{rho_sp}.
Finally, we arrive at a self-consistent set of equations for the volume fractions
and the field values, consisting of  
the constraints 
and 
 \begin{align}
 \frac{\hat\tau^{(k)}}{\varrho_0}  
 & = \sum_l \tilde q_l \frac{\tilde p_l}{z_l}  
 \exp\left\{ \frac{L\chi}{4}\left( 2 \tilde q_l \frac{\hat\tau^{(k)}}{\varrho_0} - \left( \frac{\hat\tau^{(k)}}{\varrho_0} \right)^2 \right) \right\},
\nonumber\\
& \mbox{with 
the component partition functions }
\label{eq-set-hom}\\ 
z_l &= \sum_{k'} v^{k'} \exp\left\{ 
\frac{L\chi}{4}\left( 
2 \tilde q_l \frac{\hat\tau^{(k')}}{\varrho_0} - \left( \frac{\hat\tau^{(k')}}{\varrho_0} \right)^2 
\right) 
\right\}
\nonumber
\end{align}
and the $\{\hat\omega_l^{(k)}\}$ determined implicitly.

For symmetric random triblock copolymers,
the ansatz from eq.~\xref{2phase-ans} of \emph{two\/} homogeneous phases 
yields the amplitude $\hat\tau_{\text{h}}$ given in eq.~\xref{sigma_h}
 and,
choosing $\Lambda_1 = -1$, the free-energy density $f_{\text{h}}$ from eq.~\xref{fh_multiph}.

\section{\label{app-min-cond}Equilibrium conditions at $\lambda >
  \lambda_c$}

The parameter vector $\left[ v^{(2)}, n_2, n_3 \right]$ 
of the fractionated lamellar phase 
at $\lambda > \lambda_c$ must be determined as a zero of the gradient vector
$\nabla_{\left(v^{(2)},\, n_2,\, n_3 \right)} f_{\text{frac}}$ with components
\begin{align}
\nonumber
\lefteqn{ 
\Biggl( \Biggr.
f_{\text{m}}^{(2)} - f_{\text{h}}^{(1)} + \frac{n_2 - p_2 +
n_3 - p_3}{v^{(2)}} \frac{\partial
f_{\text{h}}^{(1)}}{\partial n_2} 
}
\\ &  \mbox{} 
+ \sum_{\nu=2}^3 
n_{\nu} \ln 
\frac{n_{\nu} \left( 1 -p_2 -p_3 -v^{(2)}\left(1-n_2-n_3\right) \right)}{
\left(p_{\nu}-v^{(2)} n_{\nu}\right)\left(1-n_2-n_3\right)}
\nonumber\\  &\mbox{} 
+ \ln 
\frac{\left( 1-n_2-n_3\right)\left(1-v^{(2)}\right)}{1-p_2-p_3-v^{(2)}\left(1-n_2-n_3\right)},
\nonumber
\end{align}
\begin{align}
\label{min-cond-ffrac}
\lefteqn{  \frac{ \partial f_{\text{m}}^{(2)} }{\partial n_2} 
	+ \frac{ 1 - v^{(2)}}{v^{(2)}} \frac{\partial
          f_{\text{h}}^{(1)}}{\partial n_2} 
} 
\\  & \mbox{} 
+ \ln 
\frac{ n_2 \left( 1 -p_2 -p_3 -v^{(2)} \left(1-n_2-n_3\right) \right) }{
\left( p_2-v^{(2)} n_2 \right)\left(1-n_2-n_3\right)},
\nonumber
\end{align}
\begin{align}
\nonumber
\lefteqn{  \frac{ \partial f_{\text{m}}^{(2)} }{\partial n_3} 
	+ \frac{ 1 - v^{(2)}}{v^{(2)}} \frac{ \partial f_{\text{h}}^{(1)} }{\partial n_2} 
}
\\   & \mbox{} 
+ \ln 
\frac{n_3\left( 1-p_2-p_3-v^{(2)}\left(1-n_2-n_3\right) \right)}{
\left(p_3-v^{(2)} n_3\right)\left(1-n_2-n_3\right)}
\Biggl. \Biggr)
\nonumber
\end{align}
Here, $p_{\nu}$ denote the constant $\lambda$-defined concentrations $p_{\nu}(\lambda)$, 
whereas $n_{\nu}$ are the variable concentrations in the fractionated state.
The following relations between the partial derivatives of $f_{\text{h}}^{(1)}(v^{(2)}, n_2, n_3)$ were
inserted:
\begin{subequations}
\begin{align}
\frac{\partial f_{\text{h}}^{(1)}}{\partial n_3} & =  \frac{\partial
f_{\text{h}}^{(1)}}{\partial n_2}\\
\mbox{and }\,\frac{\partial f_{\text{h}}^{(1)}}{\partial v^{(2)}} & =  
\frac{ n_2 - p_2 + n_3 - p_3 }{ v^{(2)}\left( 1-v^{(2)} \right) } 
\frac{\partial f_{\text{h}}^{(1)}}{\partial n_2}.
\end{align}
\end{subequations}  
The fact that $f_{\text{h}}^{(1)}$ depends on the concentration
$(n_2 + n_3)$ only, 
simplifies the eq.~system~\xref{min-cond-ffrac}
for $v^{(2)}=\text{const}$ (e.g., in computing the three-phase boundaries; 
cf.\ sec.~\ref{subsec-minv01}).
From eqs.\ \xref{S-p-nu}, \xref{fmps1_min}, \xref{fh_multiph}, and \xref{constraint-p-lambda} 
one reads off the derivatives of $f_{\text h}^{(1)}$ and $f_{\text{m}}^{(2)}$ 
as functions of $v^{(2)}$, $n_2$, $n_3$:
\begin{align}
\label{dfmps2dn_nu}
\frac{\partial f_{\text{m}}^{(2)}}{\partial n_{\nu}}  = 
\Biggl\{
&\frac{ 2\chi \left(
S_{\nu}\left( k_{\text{opt}}^2 
\right) - D \left(3M, k_{\text{opt}}^2 
\right)
    \right) }{ 
\chi S\left( k_{\text{opt}}^2 
\right) - 2L } \Biggr. \\
& \mbox{} + \Biggl. \frac{16}{81} \frac{ 9 m_2^{(2)} 
+ 5 }{\left(m_2^{(2)} 
\right)^2 + m_4^{(2)} 
} 
\Biggr\}f_{\text{m}}^{(2)}, \quad\nu = 2,3, \nonumber
\end{align}
where $k_{\text{opt}}^2$, $m_{2}^{(2)}$, and $m_4^{(2)}$ are functions of 
$n_2$, $n_3$, and
\be 
\frac{\partial f_{\text{h}}^{(1)}}{\partial n_2} = - \frac{v^{(2)}}{1-v^{(2)}}
\ln\left\{ 4 \left( \cosh \frac{L\chi\hat\tau_{\text{h}}^{(1)} 
}{6\varrho_0}
\right)^2 - 3 \right\},
\ee
with the amplitude $\hat\tau_{\text{h}}^{(1)}/\varrho_0$ 
(eq.~\xref{sigma_h} with 
$p_1 = n_{1}^{(1)}$) 
expressed as a function of $v^{(2)}$, $n_2$, $n_3$ via eq.~\xref{constraint-p-lambda}.

At a given point $(\lambda, L\chi)$ of the phase space, the allowed domain $\cal{V}$ 
for the variables $v^{(2)}$, $n_2$, $n_3$ is
\begin{align} 
{\cal{V}}\mathrel{\mathop:}= \Biggl\{ 
 &\left( 
v^{(2)}\in[0,1],\, 
n_2\in\left[ 0, \min\left( 1, \frac{p_2}{v^{(2)}} \right)
\right],\,
\right.\nonumber\\
& \left. 
n_3\in\left[0, \min\left( 1-n_2, \frac{p_3}{v^{(2)}}  \right) \right]
\right):
\label{def-set} 
\\
& k_{\text{opt}}(n_2, n_3) > 0 \quad\text{and}\quad L\chi_{\text{m}}(n_2, n_3) \leq L\chi 
\Biggr\}.
\nonumber
\end{align}

\section{\label{numerical}Numerical solution of the equilibrium conditions for the fractionation free energy\label{num-sol}} 
In order to locate the zeros of the system~\xref{min-cond-ffrac}, 
which correspond to a minimum of the fractionation free energy, 
we employ a Newton-type algorithm using the following steps 
(exemplified for 
the fractionated lamellar phase):
\begin{enumerate}
\item At a given $(\lambda, \chi)$, guess start parameter vector 
$\bm x_0 \mathrel{\mathop:}= \left[ v^{(2)}_0, n_{2,0}, n_{3,0} \right]^T(\lambda, \chi)$. 
\label{startval}
The sensitivity regarding the start vector impedes completely automatized
scans in the $\lambda$-$\chi$ plane.
\item Iteratively, apply Newton scheme
\be 
\bm x_1 = \bm x_0 - H^{-1}\left(\bm x_0\right).\nabla
f_{\text{frac}}\left(\bm x_0\right),
\ee  
with 
$H$ the Hessian of the system
\xref{min-cond-ffrac}.
\item Stop if either the desired relative precision 
$\epsilon \mathrel{\mathop:}= 
\frac{\left|\bm x_1 -\bm x_0\right|}{\left|\bm x_0\right|}$ 
or a given maximal number of iterations has been reached.
In the latter case, and if $H$ gets singular during the iteration,
restart from step \ref{startval}.
\item To ensure that $f_{\text{frac}}\left(\bm x_1\right)$ is a minimum, check $H$ for positive definiteness, i.e.\ calculate its eigenvalues.
\item From the minimum concentrations $n_2$, $n_3$, calculate $k_{\text{opt}}(n_2, n_3)$ and 
$L\chi_{\text{m}}(n_2, n_3)$ 
(the continuous microphase transition that would occur in
an independent, global sequence distribution equal to the fractionated one)
and ensure the result vector to comply with 
eq.~\xref{def-set}.
\end{enumerate}

Convergence, especially while approaching the multicritical point
$(\lambda_c, L\chi_c)$, can
be achieved only for start vectors very close to the actual
solution. Therefore, proceeding on a three-phase boundary line (see section \ref{subsec-minv01})
toward $(\lambda_c, L\chi_c)$, we use the solution at one value of $\lambda$
as the start vector for the adjacent $\lambda$. 
The resolution for $\lambda$ is chosen 
between $5\times10^{-4}$ far from $\lambda_c$ and $10^{-5}$ near $\lambda_c$, 
and between $10^{-3}$ and $10^{-4}$ for $L\chi$.
In the vicinity of $(\lambda_c, L\chi_c)$, 
entries of the start vector have to be even closer to the actual solution and are obtained 
by extrapolating solutions on the boundary line.
Finally, the result vector is calculated with a relative precision
$\epsilon = 10^{-12}$ of its modulus.
Uniqueness of solutions of the nonlinear equation systems mentioned in 
secs.~\ref{sec-f-lam}--\ref{subsec-minv01} cannot be proven rigorously. 
However, we are sure not to miss transition lines to three-phase coexistence at lower $L\chi$, 
since at each $\lambda$, we start to scan the domain of definition eq.~\xref{def-set} 
with the $\lambda$-defined concentrations.


\bibliography{rand_copolys_PRE}
\end{document}